 \documentclass[aps,pra,reprint,longbibliography,10pt]{revtex4-1}

\usepackage{graphicx,float}
\usepackage{microtype}
\usepackage[svgnames]{xcolor}
\usepackage{hyperref}
\hypersetup{colorlinks,linkcolor={DarkBlue},citecolor={DarkBlue},urlcolor={DarkBlue}}
\usepackage[caption=false]{subfig}

\usepackage{amsmath}
\usepackage{mathtools}
\usepackage{amssymb,bm}
\usepackage{dsfont}
\usepackage[capitalize]{cleveref}
\usepackage{braket}
\usepackage{siunitx}
\usepackage[makeroom]{cancel}

\newcommand\defeq{\mathrel{\overset{\makebox[0pt]{\mbox{\normalfont\tiny\sffamily def}}}{=}}}
\newcommand{\e}{\operatorname{e}}

\newcommand{\tr}{\operatorname{tr}}

\begin{document}

\title{Optimal modular architectures for universal linear optics}
\author{Shreya P.~Kumar}
\email{shreya.pkumar@gmail.com}
\affiliation{Xanadu Quantum Technologies, 777 Bay Street,Toronto ON, M5G 2C8, Canada,}
\author{Ish Dhand}
\email{ishdhand@gmail.com}
\affiliation{Xanadu Quantum Technologies, 777 Bay Street,Toronto ON, M5G 2C8, Canada,}
\date{\today}
\begin{abstract}
We present modular and optimal architectures for implementing arbitrary discrete unitary transformations on light.
These architectures are based on systematically combining smaller $M$-mode linear optical interferometers together to implement a larger $N$-mode transformation.
Thus this work enables the implementation of large linear optical transformations using smaller modules that act on the spatial or the internal degrees of freedom of light such as polarization, time or orbital angular momentum.
The architectures lead to a rectangular gate structure, which is optimal in the sense that realizing arbitrary transformations on these architectures needs a minimal number of optical elements and minimal circuit depth.
Moreover, the rectangular structure ensures that each the different optical modes incur balanced optical losses, so the architectures promise substantially enhanced process fidelities as compared to existing schemes. 
\end{abstract}
\maketitle

\section{Introduction}
Linear optics is a promising route to attaining quantum computational advantage via boson sampling~\cite{Aaronson2013}, Gaussian boson sampling~\cite{Hamilton2017,Bromley2019} and quantum simulations of vibronic spectra \cite{Huh2015}, and to universal quantum computation~\cite{Rudolph2017}.
Obtaining a quantum advantage in linear optics could require scaling up to a large number of optical modes.
Consider as example boson sampling, in which beating recent classical algorithms requires tens of indistinguishable photons in several hundreds of modes~\cite{Neville2017,Clifford2018}.

Current implementations of linear optics rely on integrated photonic chips~\cite{Carolan2015,Harris2016,Flamini2019}, which allow for fast, low-loss and stable action on the spatial modes of light.
However, scaling up to a large number $N$ of modes with current methods would require integrating $N^{2}$ optical components on a single chip.
The large on-chip area required to integrate these components and the corresponding components for classical control and processing could impede scaling up to many spatial modes on photonic chips.
Thus, current photonic chips are typically limited in size by technological factors.

Alternative approaches for scaling up to higher numbers of modes involve exploiting the temporal modes of light in a single spatial mode~\cite{Motes2014,Takeda2017,Qi2018}.
These approaches promise unbounded scalability but they impose the requirements of firstly low-loss and stable optical delay lines and secondly fast reconfigurable optical elements.
These requirements could pose a daunting challenge to these temporal architectures.

To overcome these challenges to the scalability of linear optics, we present optimal modular architectures for realizing unitary transformations on light. 
The architectures are \textit{modular} as they allow combining multiple small $M$-mode optical interferometers to realize a special unitary transformation $\text{SU}(N)$ on a large number $N>M$ of modes.
In this aspect, the architectures resemble those that were introduced in Refs.~\cite{Dhand2015,Su2019a} with the motivation of realizing unitary transformations on the combined spatial and internal modes of light.
These architectures and the ones we introduce can realize an $\text{SU}(N)$ transformation using multiple linear optical \textit{modules}, each acting only on $M<N$, modes of light.
Examples of such modules include integrated photonic chips that implement unitary transformations on $M$ spatial modes~\cite{Carolan2015,Harris2016,Flamini2019}; optical loops acting on $M$ time bins~\cite{Motes2014,Takeda2017,Qi2018}; and waveplates acting on $M = 2$ polarization modes~\cite{Simon1989,Simon1990}.
Thus these architectures enable scaling up to larger total numbers of modes by combining multiple such modules together.

Moreover, the architectures presented in this work are \textit{optimal} in the sense that using these to realize arbitrary transformations needs a minimal number of required optical elements and minimal optical circuit depth.
The architectures also promise significant robustness against optical losses, which are the crucial imperfection in linear optics interferometers.
The low circuit depth and higher robustness of the architectures are a result of their rectangular structure similar to that of the Clements~\textit{et al.}\ architecture~\cite{Clements2016} rather than a triangular structure similar to that of the Reck \textit{et al.} architecture~\cite{Reck1994}.
This rectangular structure is advantageous as it firstly reduces the optical depth of the circuit as compared to a triangular structure by a factor of two thereby reducing the maximum loss acting on any of the modes. 
Secondly, it leads to each mode traversing roughly the same number of optical elements and thus incurring similar optical losses as the other modes.
This balanced structure leads to improved robustness to losses, which is especially advantageous if the implemented protocol exploits post-selection of the measured light, for instance in boson sampling~\cite{Aaronson2013}.
The rectangular structure of the architectures is different from that of Refs.~\cite{Dhand2015,Su2019a}, which have a triangular structure that is more analogous to that of the Reck~\textit{et al.}\ decomposition.
Thus, our architectures combine the advantages of modularity with those of robustness and improved circuit depth.

This article is structured as follows.
\cref{Sec:Background} reviews relevant notation.
Next, the two optimal modular architectures are presented in \cref{Sec:Architecture}, providing a detailed description of the two architectures in terms of decompositions and implementations.
This forms of presentation is continued in \cref{Appendix:Review}, which provides a unified picture of all existing architectures of linear optics clarifying the involved decompositions and the corresponding implementations.
\cref{Sec:Efficiency} presents a cost analysis in terms of number of optical elements and circuit depth and provides evidence of enhanced fidelity as compared to other modular architectures. 

\section{Background on linear optics architectures: decompositions and implementations\label{Sec:Background}}
Linear optics architectures implement discrete unitary transformations on the spatial or internal degrees of freedom of light.
More specifically, a linear optics architecture is a set of rules that describes how different optical elements can be combined to implement a desired optical transformation. 
Current architectures comprise two steps: decomposition and implementations. 
The first step is to decompose or factorize a given $\text{SU}(N)$ into smaller $\text{U}(M)$ transformations, typically for the case of $\text{U}(2)$ transformations.
The implementation step, for $M=2$ involves identifying the obtained $\text{U}(2)$ transformations by beam-splitters acting either between different spatial modes of light or between different temporal modes connected via optical delay lines~\cite{Motes2014,Takeda2017,Qi2018}.
For $M\ge 2$, the obtained $\text{U}(M)$ transformations could be used to implement transformations on the combined spatial and internal degrees of freedom of light such as polarization and orbital angular momentum or on the combined temporal and spatial modes of light.
Before presenting new architectures, let us first review some basic definitions.

\textit{Decompositions.\textemdash}
The first step of linear optics architectures are decompositions, which receive as input a special unitary matrix $U \in \text{SU}(N)$.
The given $U$ describes a linear optical transformation that maps the bosonic annihilation and creation operators $a_i, a_i^{\dagger}$ according to 
\begin{equation}
a_i \to a^{\prime}_i = \sum_{j=1}^{N} U_{ij} a_j,
\end{equation}
and similarly for the Hermitian conjugates.
Decomposition algorithms return sequences of smaller unitary transformations, such that these smaller transformations can be implemented straightforwardly using optical elements.

Decompositions rely on \textit{nulling} the entries of the matrix $U$ by multiplying it with simpler matrices in a manner analogous to Givens rotations~\cite{Wikipedia2019b}.
The nulling is typically performed using matrices of the form ${T}_{mn}(\theta,\phi)$, which differs from the $N\times N$ identity matrix $\mathds{1}_{N}$ only at the entries at the intersection of the $m$-th and $n$-th rows and columns.
These elements are set to
\begin{align}
	\begin{pmatrix*}[r]
	\e^{i\phi}\,\cos{\theta} & -\sin{\theta} \\
	\e^{i\phi}\,\sin{\theta} & \cos{\theta}
	\end{pmatrix*}.
\end{align}
That is, 
\begin{align}
T_{mn}(\theta, \phi) \defeq
\begin{pmatrix*}[r]
1 & & & & & & & \\
& \ddots& & & & & & \\
& & e^{i\phi} \cos \theta&& \hspace*{-4.5pt} - \sin \theta & & & \\
& & & \ddots & & & & \\
& & e^{i\phi} \sin \theta && \hspace*{4pt} \cos \theta & & & \\
& & & & & & \ddots & \\
& & & & & & & 1
\end{pmatrix*},
\end{align}
where the diagonal dots represent unity elements and the elements that are not shown are zero.
Note that the inverse of the $T_{mn}$ matrices is given by
\begin{align}
T_{mn}^{-1}(\theta, \phi) =
\begin{pmatrix*}[r]
1 & & & & & & & \\
& \ddots& & & & & & \\
& & e^{-i\phi} \cos \theta&& \hspace*{-4.5pt} e^{-i\phi} \sin \theta & & & \\
& & & \ddots & & & & \\
& & -\sin \theta && \hspace*{4pt} \cos \theta & & & \\
& & & & & & \ddots & \\
& & & & & & & 1
\end{pmatrix*}.
\end{align}
Henceforth, the arguments $\theta$ and $\phi$ are dropped to simplify the notation.

The $T_{mn}$ and $T_{mn}^{-1}$ matrices can be multiplied by given matrices to null their elements. 
For example, multiplying a given matrix with $T_{mn}^{-1}$ from the right leads to a new matrix that is identical to the old one except in columns $m$ and $n$, which are now mixed together with weights given by the elements of $T_{mn}^{-1}$.
Parameters $\theta$ and $\phi$ can be chosen such that the mixing results in one element out of these two columns being nulled, i.e., becoming zero after multiplication.
Existing decompositions use a sequence of $T_{mn}$ and $T_{mn}^{-1}$ matrices to null each of the elements of a given matrix~\cite{Reck1994,Clements2016,Su2019a}.

\textit{Implementations.\textemdash}
If $M = 2$, then decompositions return sequences of $T_{mn}$ and $T_{mn}^{-1}$ matrices, as these matrices can be implemented using simple optical elements.
Specifically, ${T}_{mn}(\theta,\phi)$ matrices can be implemented as a phase-shifter effecting optical phase $\phi$ on the $m$-th mode followed by a beam-splitter of transmissivity $\cos \theta$ acting between modes labeled $m$ and $n$. 
Similarly, ${T}_{mn}^{-1}$ matrices can also be implemented by a beam-splitter followed by a phase-shifter.
The modes involved in these transformation could be either spatial modes, as described in the original Reck \textit{et al.} and Clements \textit{et al.} proposals, or those in other degrees of freedom of light such as polarization~\cite{Simon1989,Simon1990}, temporal modes~\cite{Motes2014,Brecht2015a}, or orbital angular momentum~\cite{Garcia-Escartin2011}.
For completeness, the three existing $M = 2$ architectures (namely those of Reck \textit{et al.}~\cite{Reck1994}, Clements \textit{et al.}~\cite{Clements2016} and de Guise~\textit{et al.}~\cite{Guise2018}) are reviewed in \cref{Appendix:Review}.

If $M>2$, as is the case for the decompositions presented in Refs.~\cite{Dhand2015,Su2019a}, then the decomposition algorithm returns smaller $\text{U}(M)$ matrices that factorize the given $\text{SU}(N)$ matrix.
These $\text{U}(M)$ matrices can themselves be implemented as transformations acting on some degree of freedom of light and connected together in the same or some other degrees of freedom of light.
For example, Ref.~\cite{Dhand2015} proposes implementations of $\text{U}(M)$ matrices using interferometers acting on internal degrees of freedom of light (such as polarization, time, orbital angular momentum) in a single spatial mode, and connecting multiple such interferometers in the spatial domain.
Ref.~\cite{Su2019a} proposes a hybrid spatial-temporal approach, in which each of the $\text{U}(M)$ matrices is implemented using a single reconfigurable spatial interferometer and optical loops that connect the different configurations of the interferometer in time.
Please see the appendix for a detailed summary of these two architectures.

\section{Results: Optimal modular architectures}
\label{Sec:Architecture}

The modular architectures of Refs.~\cite{Dhand2015,Su2019a} suffer from suboptimal circuit depth and low fidelities resulting from imbalanced losses because of their triangular structure.
These imbalanced losses arise from some modes passing through more optical elements than others, which leads to more loss acting on some modes than others.
Here we present two decompositions of $\text{SU}(N)$ into $\text{U}(M)$ transformations that overcome this challenge. 
These decompositions allow for modular architectures, i.e., for combining smaller interferometers of arbitrary size $M$ to effect larger transformations of size $N$.
Moreover, these new architectures afford optimal optical depth, a factor of two improvement in depth as compared to existing modular architectures~\cite{Dhand2015,Su2019a},  and balanced losses because of their rectangular structure.

\subsection{Elimination-based architecture}
Here we present the first architecture for realizing $\text{SU}(N)$ transformations using $M$-mode interferometers combined together in a rectangular structure.
The decomposition relies on grouping elements from neighboring modes together into universal $M$-mode transformations and specialized $(2M-3)$-mode residual transformations similar to those of the elimination-based procedure of Ref.~\cite{Su2019a}.
In contrast to previous architectures, the rectangular structure is obtained by systematically nulling the entries of a unitary matrix by multiplying it with $T_{mn}$ from the left and from the right.

\textit{Decomposition.\textemdash}
Consider as an illustration the case of decomposing an $\text{SU}(7)$ matrix into $\text{U}(3)$ matrices.
A general $\text{SU}(7)$ matrix $U_{7}$ is represented by
\begin{align}\renewcommand{\arraystretch}{1.3} 
\left(\begin{array}{c|c|c|c|c|c|c}
*
& {\color{DarkViolet} E_{(1,2)}^{12,r}}
& {\color{DarkViolet} E_{(2,3)}^{11,r}}
& {\color{DarkGreen} D_{(1,2)}^{10\ell}}
& {\color{DarkGreen} D_{(1,2)}^{8\ell}}
& {\color{DarkRed}A_{(5,6)}^{2,r}}
& {\color{DarkRed}A_{(6,7)}^{1,r}}
\\
\hline
&*
& {\color{DarkViolet} E_{(2,3)}^{13,r}}
& {\color{DarkGrey} F_{(2,4)}^{14,r}}
& {\color{DarkGreen} D_{(2,3)}^{9\ell}}
& C_{(2,4)}^{7\ell}
& {\color{DarkRed}A_{(6,7)}^{3,r}}
\\
\hline
&& *
& {\color{DarkGoldenrod} G_{(3,4)}^{16,r}}
& {\color{DarkGoldenrod} G_{(4,5)}^{15,r}}
& {\color{DarkBlue}B_{(3,4)}^{6\ell}}
& {\color{DarkBlue}B_{(3,4)}^{4\ell}}
\\
\hline
&&&*
& {\color{DarkGoldenrod} G_{(4,5)}^{17,r}}
& {\color{DarkSalmon} H_{(4,6)}^{18,r}}
&{\color{DarkBlue} B_{(4,5)}^{5\ell}}
\\
\hline
&&&&*
& {\color{DarkCyan} I_{(5,6)}^{20,r}}
& {\color{DarkCyan} I_{(6,7)}^{19,r}}
\\
\hline
&&&&&*
& {\color{DarkCyan} I_{(6,7)}^{21,r}}
\\
\hline
&&&&&&*
\end{array}\right),
\end{align}
the bottom off-diagonal part is omitted for simplicity and, in general, the elements are complex-valued, and the elements labelled by different alphabets are given different color for ease of identification.
Specifically, the matrix elements with subscripts $(m,n)$ are nulled systematically in the order of their superscripts using $T_{mn}$ matrices.
The $\ell$ or $r$ symbols in the superscript respectively indicate whether the element is nulled by multiplying it with $T_{mn}$ matrices from the left or $T_{mn}^{-1}$ matrices from the right.

\begin{figure}
\subfloat[\label{Fig:SU7}]{
\includegraphics[width = 0.47\textwidth]{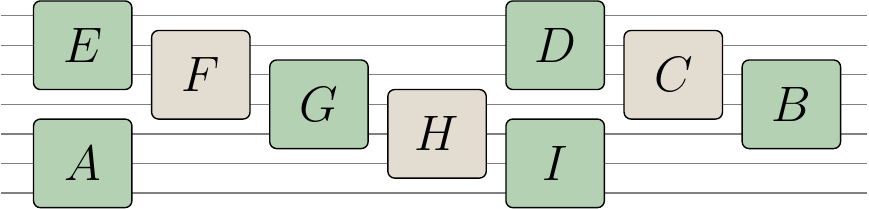}}\\
\subfloat[\label{Fig:EliminationImplement}]{
\includegraphics[width = 0.4\textwidth]{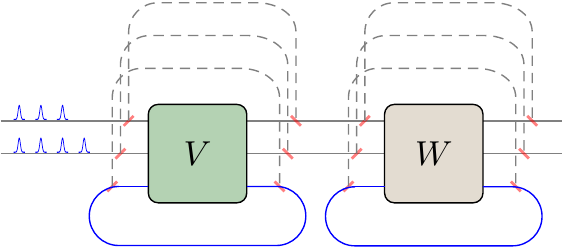}}
\caption{
(a.) Elimination-based optimal modular architectures for realizing an $\text{SU}(N)$ matrix using universal $\text{U}(M)$ and residual $\text{U}(2M-3)$ interferometers for $N, M = 7,3$. 
The green and brown boxes represent universal and residual interferometers respectively.
Note that some of the residual interferometers (E.g., $C$) are reversed as compared to those of \cite{Su2019a}.
(b.) A basic unit for hybrid spatial-temporal implementation of the circuit of (a.). 
The green and brown boxes represent spatial interferometers that implement the universal and residual unitary matrices.
The red diagonal lines represent fast switches, i.e., reconfigurable beam-splitters that toggle between perfect reflectivity and perfect transmissivity.
The blue lines are long  delay lines that implement a time delay equal to the spacing between the input pulses.
The dashed grey lines bypass the possibly lossy spatial green and brown interferometers and leave the light unchanged otherwise. 
}
\end{figure}

Thus, the decomposition begins by nulling the first three element by multiplication from the right:
\begin{alignat*}{2}
 && &U_{7}\\
\to~&& & {U_{7}}({T}_{67} {T}_{56} {T}_{67})^{-1}.
\end{alignat*}
This nulling is possible because the elements of $A$ are within a triangular block, and a nulling order similar to that of the Reck \textit{et al.} decomposition can be employed.
Then the factors in the parenthesis are combinations that can be grouped together into $\text{U}(3)$ matrices acting on three adjacent rows and leaving the other rows unchanged.
\begin{align*}
\to {U_{7}}A_{5\dots 7}^{-1}
\end{align*}
The next step involves a multiplication from the left according to 
\begin{alignat*}{2}
=~&& & {U_{7}}A_{5\dots 7}^{-1}\\
\to~ && ({T}_{34} {T}_{45} {T}_{34})&{U_{7}}
A_{5\dots 7}^{-1}\\
=~ && B_{3\dots 5}^{-1}&{U_{7}} A_{5\dots 7}^{-1}.
\end{alignat*}
This alternation between nulling from the left and nulling from the right is motivated by the Clements \textit{et al.} decomposition and is responsible for the rectangular structure of the circuit.
Proceeding along these lines and nulling all the elements gives 
\begin{equation*}
D_{1\dots 3}^{-1} C_{2\dots 4}^{-1} B_{3\dots 5}^{-1} {U_{7}} A_{5\dots 7}^{-1} E_{1\dots 3}^{-1} F_{2 \dots 4}^{-1} G_{3 \dots 5}^{-1} H_{4 \dots 6}^{-1} I_{5\dots 7}^{-1} = \mathds{D}_{7}^{\prime}
\end{equation*}
where $\mathds{D}_{7}^{\prime}$ is obtained by nulling the elements above the diagonal of $U_{7}$.
 $\mathds{D}_{7}^{\prime}$ is a diagonal matrix because any unitary matrix that is  lower triangular is diagonal. 
Let us move the matrices $A$ through $I$ to the right hand side and move the phases to the end of the circuit to obtain
\begin{align}
\begin{split}
{U_{7}}  = & \mathds{D}_{7}
B_{3\dots 5}
C_{2\dots 4}
D_{1\dots 3}\nonumber\\
& I_{5\dots 7}
H_{4 \dots 6} 
G_{3 \dots 5} 
F_{2 \dots 4} 
E_{1\dots 3} 
A_{5\dots 7},
\label{Eq:Decomposition}
\end{split} 
\end{align}
which completes the decomposition. 
The diagonal phase matrix is absorbed into the other matrices.
Thus, the given $\text{SU}(7)$ matrix is decomposed into six $\text{U}(3)$ matrices denoted $A, B, D, E, G, I$, three residual matrices $C, F, H$.
As in~\cite{Su2019a}, for $N = k(M-1)+1$ for positive integer $k$, any given ${U} \in \text{SU}(N)$ can be decomposed into $k(k+1)/2$ universal $\tilde V\in\text{U}(M)$ matrices and $k(k-1)/2$ residual matrices $\tilde W$.
The resulting structure is depicted in~\cref{Fig:SU7}.

\textit{Implementation.\textemdash}
The decomposition can be used as a basis of a variety of implementations, such as purely spatial architecture or hybrid architectures combining different degrees of freedom.
A purely spatial implementation of say $N = 7$ and $M = 3$ could allow for implementing an $\text{SU}(7)$ transformation by combining two kinds of three-mode interferometers (tritters): firstly universal twitters that implement arbitrary $U(3)$ transformations and secondly a residual tritter that mixes the first and third mode and leaves the other unchanged. 
As compared to a triangular architecture obtained from the elimination-based decomposition of Ref.~\cite{Su2019a}, a spatial implementation of the current, rectangular, architecture could lead to half the circuit depth and balanced losses.

Another possibility is a hybrid \textit{spatial-temporal} implementation, which uses reconfigurable spatial interferometers (chips), some of whose output ports are fed back into some input ports via optical delay lines.
Such an implementation requires one reconfigurable $M$-mode interferometer and another interferometer effecting the $2M-3$ residual unitary matrix, in order to perform the full transformation on a total of $N = k(M-1) + 1$ pulses of light, $k$ in each spatial mode similar to that proposed in~\cite{Su2019a}.
These interferometers can be realized using the optimal spatial implementation of Clements \textit{et al.}~\cite{Clements2016}.
The basic building block of such an approach is presented in \cref{Fig:EliminationImplement}.
Basic units such as these can either be lined up in series in a chain-loop setting~\cite{Qi2018}, or a single one can be reused by feeding its output back into its inputs in a double loop setting~\cite{Motes2014}.

The hybrid spatial-temporal implementation can be used as follows.
For concreteness, we consider $N = 7$ and $M = 3$, so $k = 3$. 
\begin{itemize}
\item
$N = 7$ pulses are impinged at the $M-1 = 2$ input ports of the chips such that the delay between these pulses is the same as the length of the delay lines.
These pulses are split into $M-1 = 2$ groups, one corresponding to each open input port of the interferometer. 
Each of the spatial modes leading to the input ports carries $k = 3$ pulses, except the last mode, which will carry $k+1 = 4$ pulses. 
\item
Each pass of the pulses through the $V$ or $W$ interferometers implements a \textit{layer} of universal or residual matrices.
For the case of the optical circuit depicted in \cref{Fig:SU7}, the layers of universal interferometer are the sets of gates $\{E\}$, $\{A, G, D\}$ and $\{I,B\}$, and the layers of residual interferometers are $\{F\}$ and $\{H,C\}$.
The layers that have fewer than $k$ universal interferometers or fewer than $k-1$ residual interferometers are padded with identity interferometers.
Thus, for the current case of \cref{Fig:SU7}, the universal layers become $\{\mathds{1},\mathds{1},E\}$, $\{A, G, D\}$ and $\{I,B,\mathds{1},\}$, and the layers of residual interferometers become $\{\mathds{1}, F\}$ and $\{H,C\}$.
\item
The operation of the optical device proceeds as follows.
First, a single pulse impinges on $V$ interferometer at the $M-1$-th input port and is directed into the delay line (i.e., the $M$-th output port) by suitably configuring the $V$ interferometer.
This pulse traverses the delay line and eventually arrives at the $M$-th input port of $V$. 
Simultaneous to the arrival of the above-mentioned cycling pulse, $M-1$ pulses impinge the first $M-1$ inputs of $V$.
\item
Then the first $M$-mode unitary of the first layer is implemented by configuring $V$ to enact this transformation.
In this case, the unitary is $\mathds{1}$.
\item
After this action, the pulses exiting the first $M-1$ output ports proceed to the next interferometer while the last output pulse is directed into the delay loop.
The interferometer is then configured to implement the next transformation of the layer, which in the example is also $\mathds{1}$.
This last pulse will interact at the linear interferometer with another $M-1$ pulses that arrive after an interval $\tau$.
This process is continued and the interferometer is configured to enact the remaining transformations of the layer one after another.
\item
A similar configuration is performed for the second interferometer $W$, which is configured repeatedly to enact the transformations of the first layer.
\item
Multiple layers can be implemented in either of two ways: (a.) by coupling a single device (of \cref{Fig:EliminationImplement}) comprising $V$ and $W$ matrices in back to itself using long optical delay lines as proposed in Ref.~\cite{Motes2015}, or (b.) by placing multiple such devices one after another in series, i.e., in a `chain-loop' configuration proposed in Ref.~\cite{Qi2018}.  
\end{itemize}

These newly-introduced identity matrices to pad the layers can be implemented by the usual reconfigurable interferometers, but such an implementation is suboptimal in terms of the incurred losses as spatial interferometers are typically lossy.
Nonetheless, if low-loss identity interferometers are available in addition to possibly lossy reconfigurable chips, then the architecture can exploit these and provides optimal incurred losses.
In this case, then the identity operations are implemented by switching the light to these low-loss interferometers, and the remaining operations are implemented as usual by lossy reconfigurable chips.
Thus, each pulse passes through the minimum number of lossy interferometers, i.e., half as many as the proposal of Ref.~\cite{Su2019a}.
Experimentally, these low-loss non-reconfigurable identity interferometers could be realized, for example, using low-loss fiber lines or additional spatial layers of the integrated chips, in which low-loss waveguides have been written.
\cref{Fig:EliminationImplement} depicts the low-loss identity operations as dashed lines that are connected to the rest of the device through switches (depicted as short diagonal lines).

\subsection{Architecture based on cosine-sine decomposition}
Here we present a rectangular modular architecture based on the cosine-sine decomposition (CSD), which we recap before presenting the decomposition. 

\textit{Background on CSD.\textemdash}
The CSD~\cite{Stewart1977,Sutton2009} factorizes a unitary matrix into three unitary matrices in a manner similar to singular value decomposition~\cite{Wikipedia2019a}.
Specifically, consider a given $(m+n)\times (m+n)$ unitary matrix $U_{m+n}$ and given integers $m,n$.
The CSD finds unitary matrices $\mathds{L}_{m+n},\mathds{S}_{m+n},\mathds{R}_{m+n}$, that factorize $U_{m+n}$ according to
\begin{equation}
U_{m+n}=
\left\{
\begin{array}{lr}
\mathds{L}_{m+n} \left(\mathds{S}_{2m}\oplus \mathds{1}_{n-m}\right)\mathds{R}_{m+n},&m\le n\\
\mathds{L}_{m+n} \left(\mathds{1}_{m-n}\oplus \mathds{S}_{2n}\right)\mathds{R}_{m+n},&m>n\\
\end{array}
\right.
\label{Eq:csd}
\end{equation}
where $\mathds{L}_{m+n}$ and $\mathds{R}_{m+n}$ are block-diagonal
\begin{equation}
\mathds{L}_{m+n} = 
\left(\begin{array}{c|c}
L_{m}& {0} 
\\
\hline
0 & L_{n}'
\end{array}\right),~
\mathds{R}_{m+n} = 
\left(\begin{array}{c|c}
R^{\dagger}_{m}& {0} 
\\
\hline
0 & R^{\prime\dagger}_{n}
\end{array} \right)
\end{equation}
with $m \times m$ and $n \times n$ blocks whose dimensions are denoted in the subscripts of the matrices.
Let us focus on the case of $m\le n$, which is the relevant case for the decompositions of~\cite{Dhand2015,Su2019a} and the new decomposition that is presented below.
The matrix $\mathds{S}_{2m}$ is an orthogonal \emph{cosine-sine (CS) matrix}, which comprises four diagonal blocks, i.e., $\mathds{S}_{2m}$ is in the form
\begin{align}
\mathds{S}_{2m} &\equiv \mathds{S}_{2m}(\theta_{1},\dots,\theta_{m})\nonumber\\
&\defeq\left(\arraycolsep=1pt\def\arraystretch{0.9}\begin{array}{ccc|ccc}
\cos\theta_1 & & &\sin\theta_1 & \\
& \ddots & && \ddots &\\
& & \cos\theta_m &&&\sin\theta_m 
\\
\hline
-\sin\theta_1 & & & \cos\theta_1 & & \\
& \ddots & && \ddots & \\
& & -\sin\theta_m &&&\cos\theta_m
\end{array}\right),
\label{Eq:CSMatrix}
\end{align}
where the dots represent more cosine and sine terms and the remaining entries are all zero.

\textit{Decomposition.\textemdash}
The decomposition can be used for any $M, N$ such that $N = \ell M$ for integer-valued $\ell$.
The decomposition receives as input an $\text{SU}(N)$ matrix and returns a sequence of arbitrary $\text{U}(M)$ transformations and specialized $2M$-mode CS transformations. 
 
Here, let us consider the case of $N = 12$ and $M = 3$.
The decomposition proceeds in two stages.
In the first stage, a suitably ordered nulling procedure decomposes the given matrix into $2M \times 2M$ matrices.
Consider a general unitary matrix
\begin{align}
U_{12} = 
\renewcommand{\arraystretch}{0.9} 
\left(\begin{array}{cccccccccccc}
*&{\color{DarkViolet}D}&{\color{DarkViolet}D}&{\color{DarkViolet}D}&{\color{DarkGreen}C}&{\color{DarkGreen}C}&{\color{DarkGreen}C}&{\color{DarkRed}A}&{\color{DarkRed}A}&{\color{DarkRed}A}&{\color{DarkRed}A}&{\color{DarkRed}A}\\
&*&{\color{DarkViolet}D}&{\color{DarkViolet}D}&{\color{DarkViolet}D}&{\color{DarkGreen}C}&{\color{DarkGreen}C}&{\color{DarkGreen}C}&{\color{DarkRed}A}&{\color{DarkRed}A}&{\color{DarkRed}A}&{\color{DarkRed}A}\\
&&*&{\color{DarkViolet}D}&{\color{DarkViolet}D}&{\color{DarkViolet}D}&{\color{DarkGreen}C}&{\color{DarkGreen}C}&{\color{DarkGreen}C}&{\color{DarkRed}A}&{\color{DarkRed}A}&{\color{DarkRed}A}\\
&&&*&{\color{DarkViolet}D}&{\color{DarkViolet}D}&{\color{DarkGoldenrod}E}&{\color{DarkBlue}B}&{\color{DarkBlue}B}&{\color{DarkBlue}B}&{\color{DarkRed}A}&{\color{DarkRed}A}\\
&&&&*&{\color{DarkViolet}D}&{\color{DarkGoldenrod}E}&{\color{DarkGoldenrod}E}&{\color{DarkBlue}B}&{\color{DarkBlue}B}&{\color{DarkBlue}B}&{\color{DarkRed}A}\\
&&&&&*&{\color{DarkGoldenrod}E}&{\color{DarkGoldenrod}E}&{\color{DarkGoldenrod}E}&{\color{DarkBlue}B}&{\color{DarkBlue}B}&{\color{DarkBlue}B}\\
&&&&&&*&{\color{DarkGoldenrod}E}&{\color{DarkGoldenrod}E}&{\color{DarkCyan}F}&{\color{DarkBlue}B}&{\color{DarkBlue}B}\\
&&&&&&&*&{\color{DarkGoldenrod}E}&{\color{DarkCyan}F}&{\color{DarkCyan}F}&{\color{DarkBlue}B}\\
&&&&&&&&*&{\color{DarkCyan}F}&{\color{DarkCyan}F}&{\color{DarkCyan}F}\\
&&&&&&&&&*&{\color{DarkCyan}F}&{\color{DarkCyan}F}\\
&&&&&&&&&&*&{\color{DarkCyan}F}\\
&&&&&&&&&&&*
\end{array}\right)
\end{align}
where alphabets are presented in different color for ease of identification..
This matrix is nulled in the order denoted by the alphabets in the above equation.
That is, the elements in the first group labelled $A$ are nulled by $T_{mn}^{-1}$  matrices acting from the right.
As before, this nulling is possible because the elements labelled as $A$ comprise a triangular block, so a nulling order similar to that of the Reck \textit{et al.} decomposition can be employed.
The exact order of nulling is presented in detail in~\cref{Sec:AppendixCSD}.
These $T_{mn}$ matrices are then grouped together into a single $\text{U}(6)$ transformation acting on the last six modes as 
\begin{align}
U_{12}A_{7 \dots 12}^{-1} = U_{\bar{A}},
\end{align}
where $A_{7 \dots 12}$ denotes a unitary transformation acting on modes $7$ through $12$.

Next the elements labelled by $B$ and $C$ are nulled by multiplying from the left by $T_{mn}$ matrices.
As in the elimination-based decomposition described above, this alternation between nulling from the left and nulling from the right is responsible for the rectangular structure of the circuit.
The $T_{mn}$ matrices used to null $B$ and $C$ elements are then gathered again into the $\text{U}(6)$ transformations $B_{4\dots 9}$ and $C_{1\dots 6}$ as 
\begin{align}
C_{1\dots 6}^{-1}B_{4\dots 9}^{-1}U_{12}A_{7\dots 12}^{-1} = U_{\bar{C}}.
\end{align}
Finally, the elements labelled $D$ through $F$ are nulled, again using $T_{mn}^{-1}$ matrices to obtain:
\begin{equation}
C_{1\dots6}^{-1}B_{4\dots9}^{-1}U_{12}A_{7\dots12}^{-1}
D_{1\dots6}^{-1} E_{4\dots9}^{-1} F_{7\dots12}^{-1} = \mathds{D}_{12},
\end{equation}
or equivalently 
\begin{equation}
U_{12}
   =  B_{4\dots9} C_{1\dots6} F_{7\dots12} E_{4\dots9} D_{1\dots6} A_{7\dots12},
   \label{Eq:CSDIntermediate}
\end{equation}
where the diagonal phases are absorbed into the unitary $\text{U}(6)$ transformations.

\begin{figure}
\subfloat[\label{Fig:CSDIntermediate}]{
\includegraphics[width = 0.47\textwidth]{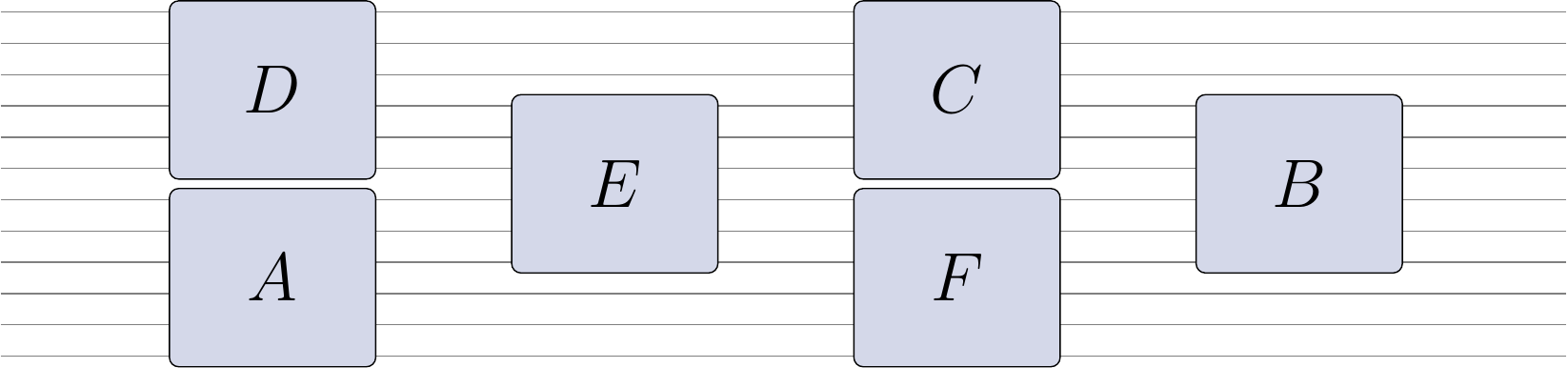}}\\
\subfloat[\label{Fig:CSDFinal}]{
\includegraphics[width = 0.47\textwidth]{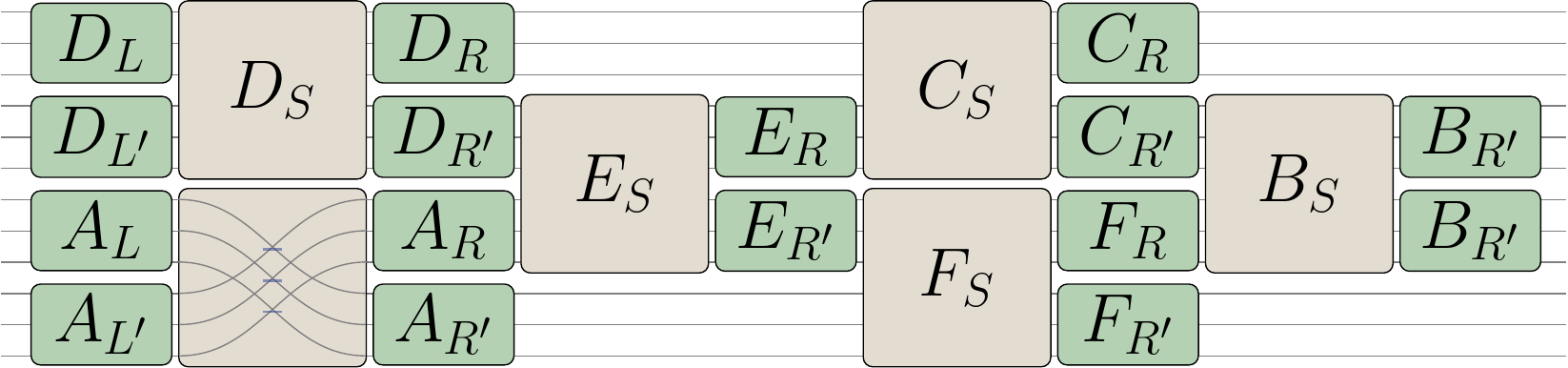}}\\
\subfloat[\label{Fig:CSDImplement}]{
\includegraphics[width = 0.47\textwidth]{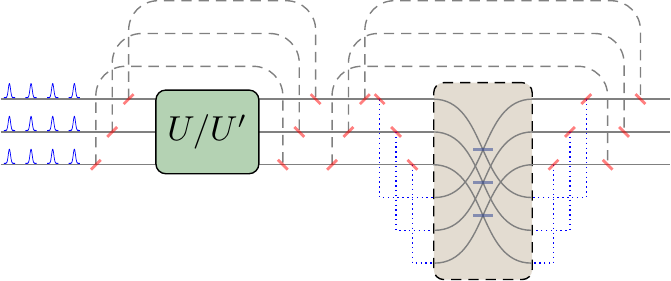}}
\caption{CSD-based modular architectures for realizing $\text{SU}(N)$ matrices.
(a.) Intermediate stage of the decomposition into $\mathrm{U}(2M)$ transformations in a rectangular structure as presented in \cref{Eq:CSDIntermediate}.
(b.) Final stage of the decomposition using universal $\text{U}(M)$ and specialized $\text{U}(2M)$ CS interferometers for $N, M = 12,3$. 
The green and brown boxes represent $M \times M$ universal unitary matrices and specialized $2M \times 2M$ CS matrices respectively.
The blue lines inside the brown box represent reconfigurable beam-splitters with arbitrary reflectivity. 
(c.) A basic unit for hybrid spatial-temporal implementation of the circuit of (b.).
\label{Fig:SU12}}
\end{figure}

At this stage, the given $12 \times 12$ matrix is decomposed into six transformations, each acting on $2M = 6$ modes.
Amongst themselves, these $\text{U}(6)$ transformations possess a rectangular structure similar to that of Clements \textit{et al.}\ as depicted in \cref{Fig:CSDIntermediate}.
However, at this stage, not all the $\text{U}(6)$ transformations are universal as some of these comprise fewer than the minimal number (six) of $T_{mn}$ parameters required for universality.
The next stage of the decomposition removes this redundancy of number of parameters.

The next stage begins by identifying that the $N$ modes are partitioned into groups of $M$ modes each.
Moreover, the $2M$-mode transformations of the first step (\cref{Fig:CSDIntermediate}) act on two nearest partitions.
In the current specific example of $M = 3$, the modes are partitioned into subsets $\{1,\dots,3| 4,\dots,6 | 7,\dots,9|10,\dots,12\}$ of three-mode elements, and the transformations obtained above act on the six-mode sets $\{1,\dots,6| 4,\dots,9| 7,\dots,12\}$.
The next step (depicted in \cref{Fig:CSDFinal}) is to decompose each of these transformations further using the CSD with the parameters $m = n = M$, which is equal to $3$ in the current case.
The action of CSD on each of the transformations leads to a sequence of two types of unitary matrices: firstly $M$-mode unitary matrices acting on individual partitions and secondly $2M$-mode CS matrices acting on neighboring partitions. 
Because the $M$-mode unitary matrices act only on individual partitions, those matrices that act sequentially on the same partition can be merged into single $\text{U}(M)$ transformation.
This merge removes the above-mentioned redundancy and reduces the circuit depth of the resulting architecture.
This completes the decomposition.

\textit{Implementation.\textemdash}
The decomposition can be used as a basis of a variety of implementations, such as purely spatial or hybrid architectures combining different degrees of freedom.
A purely spatial implementation for the instance of $N = 12$ and $M = 3$ could involve realizing a $\text{SU}(12)$ transformation by combining three-mode interferometers (tritters) and specialized six-mode CS interferometers, each of which can be realized using three beam-splitters.
Such an architecture would be useful in a situation where spatial interferometers are limited in the number of modes they can act on, but they can be connected together via low-loss interconnects.
A hybrid \textit{internal-spatial} implementation could involve combining firstly modules enacting universal unitary transformations on $M = 3$ dimensional internal degrees of freedom such as polarization, time bins, temporal modes, or orbital angular momentum, and secondly modules implementing the CS transformation as proposed in~\cite{Dhand2015}.

Finally, a hybrid spatial-temporal implementation similar to that described for the elimination based decomposition above can be employed.
The basic building block of such an implementation is presented in \cref{Fig:CSDImplement}.

In more detail a hybrid spatial-temporal implementation based on this decomposition can be used as follows.
For concreteness, we consider $N = 12$ and $M = 3$, so $\ell = M/N = 4$ and we depict this example in \cref{Fig:CSDImplement}
\begin{itemize}
\item
$N = 12$ pulses are impinged at the $M = 3$ input ports of the chips such that the delay between these pulses is the same as the length of the delay lines.
These pulses are split into $M$ groups, one corresponding to each open input port of the interferometer. 
Each of the spatial modes leading to the input ports carries $\ell = 4$ pulses.
\item
Each pass of the pulses through the $U/U^\prime$ or CS interferometers implements a layer of universal or CS matrices.
For the case of the optical circuit depicted in \cref{Fig:CSDFinal}, the layers of universal interferometer are the sets of gates $\{A_{L'}, A_{L}, D_{L} , D_{L'}\}$, $\{A_{R'}, A_{R}, D_{R'}, D_{R}\}$, $\{E_{R'}, E_{R}\}$, $\{F_{R'}, F_{R}, C_{R'}, C_{R}\}$, and $\{B_{R'}, B_{R}\}$ and the layers of CS interferometers are $\{A_{S}, D_{S}\}$, $\{E_{S}\}$, $\{F_{S}, C_{S}\}$ and $\{B_{S}\}$.
Note that $\{A_{S}\}$ is depicted as the unlabeled interferometer below the $D_{S}$ interferometer in \cref{Fig:CSDFinal}.
The layers that have fewer than $\ell$ universal interferometers or fewer than $\ell/2$ CS interferometers are padded with identity interferometers acting on $M$ modes each.
Thus we have the following universal layer after padding: $\{A_{L'}, A_{L}, D_{L} , D_{L'}\}$, $\{A_{R'}, A_{R}, D_{R'}, D_{R}\}$, $\{\mathds{1}, E_{R'}, E_{R}, \mathds{1}\}$, $\{F_{R'}, F_{R}, C_{R'}, C_{R}\}$, and $\{\mathds{1}, B_{R'}, B_{R}, \mathds{1}\}$,  and the padded CS layer: $\{A_{S}, D_{S}\}$, $\{\mathds{1}, E_{S}, \mathds{1}\}$, $\{F_{S}, C_{S}\}$ and $\{\mathds{1}, B_{S}, \mathds{1}\}$.
Notice that no more than two padding identity interferometers are required in each layer irrespective of the values of $N$ and $M$, one at the beginning and one at the end of the layer.
\item
The operation of the device proceeds as follows.
First, a set of $M$ pulses impinge on the $M$ input ports of the $U/U'$ interferometer. 
The first $M$-mode unitary of the first layer is implemented by configuring $U/U'$ to enact this transformation, which is $A_{L'}$ in this case.
\item
The interferometer then enacts the next transformation $A_{L}$ on the next $M$ pulses that impinge at the input ports. 
In a similar manner, the first interferometer is reconfigured to perform the complete first layer of transformations on the modes.
\item
After this action, the pulses exiting the output ports are directed into the CS interferometer, which is configured to implement the first CS layer on $2M$ pulses that arrive in groups of $M$ pulses at two different arrival times.
Of these $2M$ pulses, $M$ impinge directly at the $M$ input ports while the other $M$ are delayed (using blue dotted lines of \cref{Fig:CSDImplement}) to arrive at the other $M$ input ports simultaneously with the first $M$ pulses.
After the action of the CS interferometer, these first $M$ pulses exit directly whereas the next $M$ pulses are delayed again and switched into the same spatial modes as the first $M$ pulses. 
\item
By repeatedly reconfiguring the CS interferometer, the complete layer of CS matrices is implemented. 
\item
As before, multiple layers can be implemented either by coupling a single device back to itself using long  delay lines, or by placing multiple such devices one after another in series in a `chain-loop' configuration. 
\end{itemize}

As in the previous elimination-based scheme, this CS-based decomposition provides an optimal factor of two enhancement in the maximum number of interferometers that any pulse has to pass through.
Moreover, this scheme does not require access to low-loss identity interferometers to provide this enhancement.
This reduced requirement is because only a constant number (two) of padding identity interferometers is required for each layer in the decomposition.
This is in contrast to the elimination decomposition, in which the number of padding interferometers required in each layer scales linearly with $N$.
This completes the discussion of implementation details of the two architectures.

\section{Analysis of cost and robustness to losses}
\label{Sec:Efficiency}

\subsection{Cost analysis}
Here we presents a brief cost analysis of the architectures.
We show that the architectures introduced here are optimal in terms of number of optical elements required in the implementation and offer substantially enhanced circuit depth as compared to existing architectures.

Which and how many optical elements are required in an implementation depend, in general, on the degrees of freedom of light that are chosen for the implementation.
We focus first on the purely spatial implementation, i.e., one in which multiple spatial interferometers are connected together.
In this case the relevant resource is the number of required beam-splitters and phase-shifters.
In this purely spatial setting, both the CSD and the elimination-based architectures require the minimum number $N(N-1)/2$ of beam-splitters.
Similarly a minimum number $N(N-1)/2$ physical phase-shifters is required in the implementation.
Another important setting is the hybrid spatial-temporal configuration, in which the important metric is the number of passes through beam-splitters and phase-shifters as each pass leads to additional optical loss.
Here too, the architectures require the minimum number $N(N-1)/2$ of passes through beam-splitters and the minimum number $N(N-1)/2$ of passes through reconfigurable phase-shifters. 
Thus, both decompositions lead to optimal architectures in terms of required optical elements or passes through optical elements. 

Next, let us consider the optical circuit depth of the architectures in terms of the number of interferometers that each mode traverses through, universal or otherwise.
It is crucial to keep the optical depth as low as possible as higher depths lead to smaller transmissivities per photon, with the transmissivity scaling exponentially towards zero in the depth.
Consider first the CSD-based architecture introduced in Ref.~\cite{Dhand2015} for parameters $M,N$ and $\ell = N/M$.
This triangular architecture has circuit depth of $2\ell-2$ universal unitary transformations and $2\ell-3$ cosine-sine transformations that act on the light.
In contrast, our current CSD-based architecture has a circuit depth of $\ell+1$ universal and $\ell$ cosine-sine transformations.

These depth scalings are optimal for an architecture that employs $M$-mode universal and $2M$-mode CS interferometers.
To see this, we observe that the set of all the modes is partitioned into subsets $\{1,2,\dots,M|M+1,M+2,\dots,2M|\dots|N-M+1,N-M+2,\dots N\}$ of $M$ modes each.
Importantly, modes from each subset mix with those of their nearest neighboring subset at the first possible occasion via CS matrices.
This minimum depth $\ell$ is analogous to the minimum depth $N$ of the rectangular decomposition of Ref.~\cite{Clements2016}. 
Now consider universal interferometers, which act only within one of the $\ell$ subsets.
In between the action of each CS matrix, only a single universal $M$-mode matrix acts on each subset. 
Hence, this architecture is optimal in terms of the $2M$-mode CS and $M$-mode universal interferometers required to implement the given transformation.

Consider now the elimination-based architecture of Ref.~\cite{Su2019a} for $k = (N-1)/(M-1)$.
Because of its triangular structure, this architecture has a circuit depth of $2k-1$ universal interferometers and $2k-3$ residual interferometers.
The rectangular architecture based on the elimination-regrouping approach has a circuit depth of $k+1$ universal and $k$ residual interferometers.
Thus this work provides a factor two improvement in the optical depth of modular architectures for linear optics.

Similar arguments as above can show the optimality of this decomposition in terms of $M$-mode universal and $2M-3$ mode residual matrices.
Specifically, an elimination-based scheme divides the set of all modes into overlapping subsets of $M+1$ elements each as $\{1,2,\dots,M+1|M+1,M+2,\dots,2M+1|\dots|N-M,N-M+1,\dots,N\}$.
These subsets are acted upon by universal interferometers and they mix with their neighboring subsets at the first possible occasion as well.
Thus, the two architectures presented here have optimal depth in terms of numbers of interferometers that each mode traverses through.

\begin{figure}
\subfloat[\label{Fig:NumberOfModes}]{
\includegraphics[width = 0.47\textwidth]{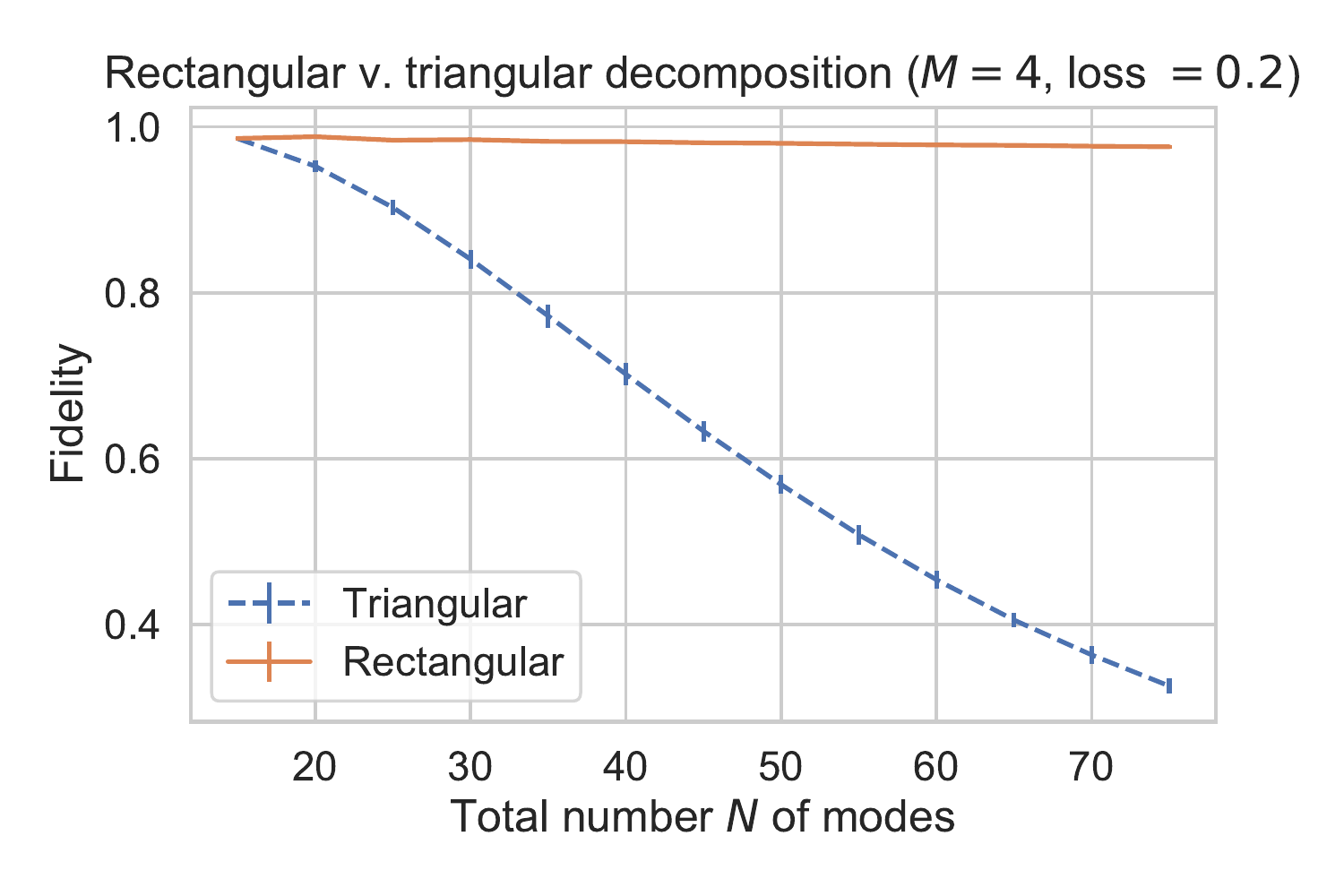}}\\
\subfloat[\label{Fig:Transmission}]{
\includegraphics[width = 0.47\textwidth]{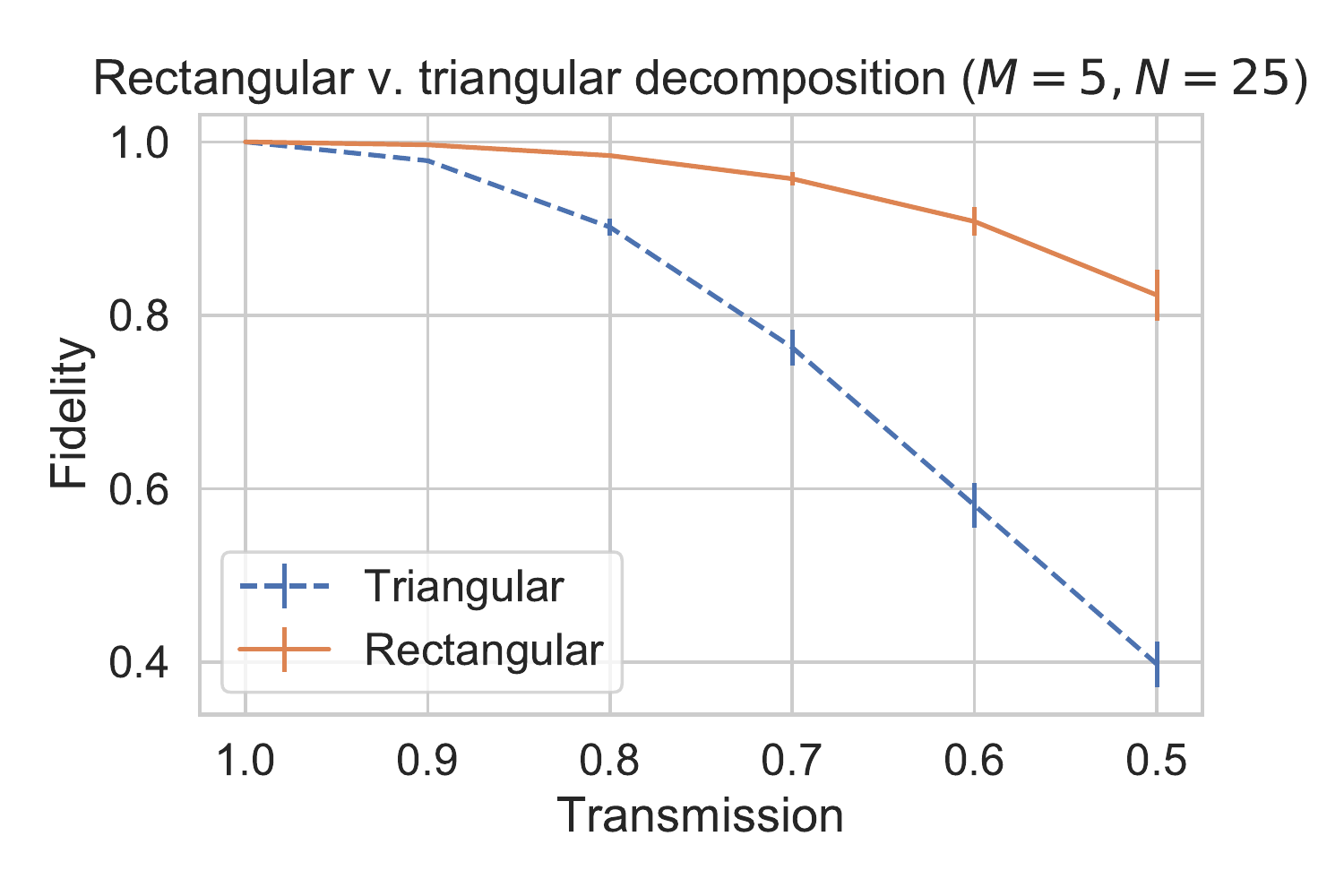}}\\
\caption{Comparison of fidelities of rectangular and triangular decompositions. 
For concreteness, we focus on the CSD based decomposition, but similar plots can be obtained for the elimination based decomposition as well.
(a) Scaling of the average fidelity as a function of the total number $N$ of modes of the system. 
Each module acts on $M = 4$ modes and adds 20\% loss, which is assumed to act uniformly on each of the modes of the module.
(b) Scaling of average fidelity with the transmission for $N = 25$ and $M = 5$.
The error bars (sometimes smaller than plot line-width) represent a spread of fidelity over two standard deviations as estimated by sampling over a hundred unitary matrices from the Haar measure.
\label{Fig:Loss}}
\end{figure}

\subsection{Loss tolerance}

In addition to the reduced circuit depth, the architectures presented here promise significantly enhanced fidelities. 
This improvement results from each of the modes passing through a similar number of optical elements because the smaller modules are combined together in a rectangular pattern.
This rectangular structure is in contrast to the triangular structure of Refs.~\cite{Dhand2015,Su2019a}, in which some of the modes pass through many more modules than others.

A comparison of the fidelities of our architectures and those of Refs.~\cite{Dhand2015,Su2019a} is plotted in \cref{Fig:Loss}.
To obtain the fidelity plots, we draw random $N \times N$ unitary matrices $U$ from the Haar measure, which are then decomposed into a sequence of $\text{U}(M)$ transformations according to the procedure described in this work.
A uniform loss is applied to each of the individual modules. 
The effective experimental unitary transformation $U_\text{lossy}$ is then obtained by multiplying the $\text{U}(M)$ transformations.
The fidelity of the lossy unitary transformation can be quantified as~\cite{Clements2016}
\begin{equation}
	F = \left| \frac{ \tr\left( U^{\dag} U_\text{lossy} \right) }{ \sqrt{N \tr \left(U^{\dag}_\text{lossy} U_\text{lossy} \right)}} \right|^2,
\end{equation}
which is insensitive to overall uniform loss.
Such a measure is relevant in settings where post-selection can be performed, in which case any overall uniform loss can be neglected.
This fidelity is then averaged over the Haar measure by sampling a hundred random unitary matrices for each data point of \cref{Fig:Loss}.

We  note that our rectangular architecture provides substantially enhanced fidelities across different values of $N$ and loss, as compared to the earlier triangular decompositions.
In summary, because of its lower circuit depth and rectangular structure, this modular architecture promises enhanced fidelities in implementing linear optical transformations.

\section{Conclusion}
In summary, we have presented modular and efficient architectures for implementing $\text{SU}(N)$ transformations using smaller modules that implement $\text{U}(M)$ transformations.
The architectures introduced here can be implemented using either an all spatial approach, or using hybrid internal spatial or spatial-temporal approaches.
Because of their rectangular structure, these architectures promise optimal circuit depth and robustness to optical losses.
This modular approach promises enhanced scalability as compared to existing non-modular approaches.
This enhanced scalability results from its lower circuit depth and balanced loss structure, which lead to higher fidelities of performing linear optical transformations.

Our architectures could be useful in implementing not only quantum information processing tasks, but also towards tasks that rely on acting linear optics transformation on many modes of classical light.
Examples of such tasks include multi-port optical switches that are currently being developed for use in data centers~\cite{Hinton2013,Cheng2018}.
Another potential use case of the procedure is towards optical neural networks, which rely on the action of linear optical interferometers on classical or quantum light~\cite{Shen2017,Tait2017}.

\acknowledgments{
We are grateful to Lukas G.~Helt and Daiqin Su for helpful comments.
}

\newpage~
\newpage

\appendix
\section{Review of existing decompositions}
\label{Appendix:Review}
For completeness, here we present a detailed exposition of existing architectures in a unified notation that is analogous to that used in the main text.
Existing architectures rely on decomposition either into two-mode unitary matrices, i.e, those introduced in Refs.~\cite{Reck1994,Clements2016,Guise2018}, or into unitary matrices acting on more than two modes, namely those of Refs.~\cite{Dhand2015,Su2019a}.
The $M=2$ architectures are presented below in Section~\ref{Sec:TwoModes} and those with $M>2$ are presented in Section~\ref{Sec:BackgroundUM}.

\subsection{Architectures based on decompositions into two-mode unitary matrices}
\label{Sec:TwoModes}
Decompositions of $\text{SU}(N)$ unitary transformations into two-mode, i.e., $\text{U}(2)$, transformations include those of Reck~\textit{et al.}~\cite{Reck1994}, Clements~\textit{et al.}~\cite{Clements2016} and of de Guise~\textit{et al.}~\cite{Guise2018}.
These decompositions enable implementations on spatial or temporal degrees of freedom of light.
Below we describe the three decompositions and provide details about relevant implementations.

\subsubsection{Triangular architecture due to Reck~\textit{et al.}}
\textit{Decomposition.\textemdash}
The Reck~\textit{et al.}\ decomposition~\cite{Reck1994} relies on nulling the elements of the given $U$ matrix.
The nulling is performed by multiplying $U$ from the right with $T_{mn}^{-1}$ matrices. 
To illustrate the order of multiplication, consider the decomposition of an $\text{SU}(5)$ matrix $U_{5}$, where the subscript of $U$ represents the number of modes that the transformation can act non-trivially on.
The first round of the decomposition nulls the last row and columns of $U_{5}$ as
\begin{equation}
U_{5} T^{-1}_{12} T^{-1}_{23} T^{-1}_{34} T^{-1}_{45} = U_{4}\oplus \mathds{D}_{1},
\label{Eq:Reck0}
\end{equation}
where $\mathds{D}_{i}$ represents an $i\times i$ diagonal unitary matrix.
The resulting $4\times 4$ unitary matrix $U_{4}$ is decomposed further into smaller and smaller matrices by multiplying with $T_{mn}^{-1}$ matrices in the order:
\begin{align}
\begin{split}
&U_{5} T^{-1}_{12} T^{-1}_{23} T^{-1}_{34} T^{-1}_{45} \\
\to~
&U_{5} T^{-1}_{12} T^{-1}_{23} T^{-1}_{34} T^{-1}_{45} 
T^{-1}_{12} T^{-1}_{23} T^{-1}_{34}\\
\to~
&U_{5} T^{-1}_{12} T^{-1}_{23} T^{-1}_{34} T^{-1}_{45}
T^{-1}_{12} T^{-1}_{23} T^{-1}_{34} 
T^{-1}_{12} T^{-1}_{23} \\
\to~
&U_{5} 
T^{-1}_{12} T^{-1}_{23} T^{-1}_{34} T^{-1}_{45}
T^{-1}_{12} T^{-1}_{23} T^{-1}_{34} 
T^{-1}_{12} T^{-1}_{23} 
T^{-1}_{12},\label{Eq:ReckLayers}
\end{split}
\end{align}
where each line represents one round of the decomposition.
Finally, these multiplications result in
\begin{equation}
U_{5} 
T^{-1}_{12} T^{-1}_{23} T^{-1}_{34} T^{-1}_{45}
T^{-1}_{12} T^{-1}_{23} T^{-1}_{34} 
T^{-1}_{12} T^{-1}_{23} 
T^{-1}_{12} 
= \mathds{D}_{5},
\label{Eq:ReckIntermediate}
\end{equation}
or equivalently 
\begin{equation}
U_{5} 
= \mathds{D}_{5} T_{12} T_{23} T_{12} T_{34} T_{23} T_{12} T_{45} T_{34} T_{23} T_{12}.
\label{Eq:Reck}
\end{equation}
The decomposition thus factorizes a given $N$-mode transformation into a sequence of $N(N-1)/2$ transformations that act on two modes each.

\textit{Implementation.\textemdash}
Reck \textit{et al.}~proposed a spatial implementation based on their decomposition~\eqref{Eq:Reck}. 
The spatial implementation returns an optical circuit that implements transformation $U$ on $N$ spatial modes of light.
This circuit comprises $N(N-1)/2$ beam-splitters and phase-shifters as basic building blocks.
Each $T_{mn}$ matrix is implemented by a phase-shifter implementing phase $\phi$ on mode $m$ followed by a beam-splitter of transmissivity $\cos\theta$ acting between modes $m$ and $n$.  
Beam-splitters acting only on neighboring modes are needed as $n = m + 1$ for each of the $T_{mn}$ matrices of \cref{Eq:Reck}.
The $\mathds{D}_{N}$ matrix in \cref{Eq:Reck} is implemented using $N$ additional phase-shifters, one acting on each mode.
The beam-splitters are arranged in a triangular structure: only a single beam-splitter acts on the last mode ($m = 5$) and $2N-3 = 7$ beam-splitters act on the second mode.
The circuit depth of the Reck~\textit{et al.}\ decomposition in terms of beam-splitters is thus $2N-3$.

The Reck~\textit{et al.}\ decomposition also provides a method to realize arbitrary transformations on $N$ temporal pulses of light as demonstrated by Motes~\textit{et al.}\ in 2014~\cite{Motes2014} and detailed in Ref.~\cite{Motes2015}.
The basic building block of the architecture is a beam-splitter with one output port feeding back into one of its input ports.
This looped beam-splitter implements a beam-splitter transformation between any two subsequent pulses of light impinged at the remaining input port if the inter-pulse spacing equals the length of the delay line. 
By using a fast reconfigurable beam-splitter, this device can act on multiple temporal modes with different beam-splitter transformations. 

The effect of the looped beam-splitter can be seen more clearly after reordering the full Reck decomposition of \cref{Eq:Reck} as
\begin{align}
U & = \, \mathds{D}_{5} T_{12} T_{23} T_{12} T_{34} T_{23} T_{12} T_{45} T_{34} T_{23} T_{12}\label{Eq:ReckBefore}\\
 & = \, \mathds{D}_{5} (T_{12} T_{23} T_{34} T_{45}) (T_{12} T_{23}  T_{34}) (T_{12} T_{23}) (T_{12})\label{Eq:ReckAfter},
\end{align}
where each of the parentheses enclose a layer of beam-splitter transformations.
Going from \cref{Eq:ReckBefore} to \cref{Eq:ReckAfter} is possible because two $T_{mn}$ matrices commute with each other if both act entirely on different modes.

If multiple equally spaced pulses are impinged at a device comprising a reconfigurable phase-shifter and a reconfigurable looped beam-splitter, then one layer of $T_{mn}$ matrices is implemented on the pulses.
Moreover, multiple such devices arranged in series can implement a composition of multiple such layers.
Thus, $N-1 = 4$ such devices arranged in series, or alternatively a single device connected to itself through a long delay line, can implement the desired $N-1$ layers of beam-splitters as returned by the full Reck~\textit{et al.}\ decomposition.

\subsubsection{Rectangular architecture due to Clements~\textit{et al.}}
\textit{Decomposition.\textemdash}
The Clements~\textit{et al.}\ decomposition~\cite{Clements2016} improves over that of Reck~\textit{et al.}\ by providing a rectangular circuit structure in which each mode is acted upon by no more than $N$ beam-splitters.
The rectangular structure is obtained by nulling the elements of the given unitary matrix systematically from the right and also from the left.
More specifically, $U_{5}$ is nulled by multiplying it with ${T}_{mn}$ matrices from the right and ${T}_{mn}^{-1}$ matrices from the left.
For instance, an $\text{SU}(5)$ matrix $U$ is converted to diagonal matrix in the order:
\begin{alignat*}{2}
 && &U_{5} T^{-1}_{12}\\
\to~ &&T_{45} T_{34} &U_{5} T^{-1}_{12}\\
\to~ &&T_{45} T_{34} &U_{5} T^{-1}_{12} T^{-1}_{34} T^{-1}_{23} T^{-1}_{12}\\
\to~ &&T_{45}T_{34}T_{23}T_{12} T_{45} T_{34} &U_{5} T^{-1}_{12} T^{-1}_{34} T^{-1}_{23} T^{-1}_{12}
\end{alignat*}
which leads to
\begin{equation}
T_{45}T_{34}T_{23}T_{12} T_{45} T_{34} U_{5} T^{-1}_{12} T^{-1}_{34} T^{-1}_{23} T^{-1}_{12} = \mathds{D}_{5}.
\end{equation}
or equivalently
\begin{equation}
U_5 = 
T^{-1}_{45}
T^{-1}_{34}
T^{-1}_{23}
T^{-1}_{12}
T^{-1}_{45}
T^{-1}_{34}
\mathds{D}_{5}
T_{12}
T_{23}
T_{34}
T_{12}.
\label{Eq:PhaseMiddle}
\end{equation}

In \cref{Eq:PhaseMiddle}, the diagonal matrix $D_5$ appears in the middle of the decomposition but these additional phases can be moved through the $T^{-1}$ matrices as follows.
New matrices $T'_{mn}$ and $\mathds{D}'_{5}$ are constructed such that $T^{-1}_{mn} \mathds{D}_{5} = \mathds{D}_{5} T_{mn}$.
For the construction of such matrices, consider operators acting on the two-mode subspace
\begin{align}
	T(\theta,\phi) = \,&
	\begin{pmatrix*}[r]
	\e^{i\phi}\,\cos{\theta} & -\sin{\theta} \\
	\e^{i\phi}\,\sin{\theta} & \cos{\theta}
	\end{pmatrix*},\\
	D(\alpha,\beta) = \,& 
	\begin{pmatrix*}[r]
	\e^{i\alpha} & 0 \\
	0 & \e^{i\beta}
	\end{pmatrix*}.
\end{align}
For these matrices, 
\begin{align}
	T^{-1}(\theta,\phi)D(\alpha,\beta) = 
	\begin{pmatrix*}[r]
	\e^{i(\alpha-\phi)}\,\cos{\theta} & \e^{i(\beta-\phi)}\,\sin{\theta} \\
	-\e^{i\alpha}\,\sin{\theta} & \e^{i\beta}\,\cos{\theta}
	\end{pmatrix*},
\end{align}
which is equal to $D'(\alpha'\beta') T(\theta',\phi')$ for
\begin{align}
\begin{split}
	\alpha' = \,& \beta - \phi + \pi\\
	\beta' = \,& \beta\\
	\theta' = \,& \theta\\
	\phi' = \,& \alpha - \beta + \pi  
\end{split}
\end{align}
Thus, we can move the phases, two at a time, through the $T$ matrices using $D'(\alpha'\beta') T(\theta',\phi') = T^{-1}(\theta,\phi)D(\alpha,\beta)$.
By moving all the phases to the left of the equation, we obtain the decomposition 
\begin{equation}
U_{5} = \mathds{D}_{5} T_{34} T_{45} T_{12} T_{23} T_{34} T_{45} T_{12} T_{23} T_{34} T_{12}.
\end{equation}
This decomposition has a rectangular structure, in which five $T$ matrices act on each of the modes except the first and last mode, on which three and two gates act respectively.

\textit{Implementation.\textemdash}
The Clements \textit{et al.} decomposition enables a spatial implementation that has optimal circuit depth. 
As in the Reck \textit{et al.} architecture, the $T_{mn}$ and $D$ matrices are implemented as beam-splitters and phase-shifters acting on different spatial modes. 
But in contrast to the Reck \textit{et al.} implementation, the first and last mode are acted upon by $\lceil (N+1)/2\rceil =3$ and $\lfloor (N-1)/2\rfloor =2$ beam-splitters and the remaining modes are acted upon by $N = 5$ beam-splitters.
Thus, the circuit depth of this architecture is $N$, which is the minimal possible depth for spatial circuits.

Furthermore, as each of the pulses traverses a similar number of optical elements, losses acting on the light are balanced.
If the losses acting on each of the mode are exactly identical, then the actual realized transformation differs from the desired unitary by a constant multiplicative factor. 
This multiplicative factor can be mitigated in postselection based linear optics protocols such as boson sampling but can pose substantial challenges to other protocols such as Gaussian boson sampling.

The Clements \textit{et al.} decomposition can also be implemented in the temporal modes of light.
The switching pattern required to implement this decomposition is seen by grouping the factors of the decomposition into layers according to
\begin{align}
U_{5} & =\, \mathds{D}_{5} T_{34} T_{45} T_{12} T_{23} T_{34} T_{45} T_{12} T_{23} T_{34} T_{12}\\
& =\, \mathds{D}_{5} (T_{34} T_{45}) (T_{12} T_{23} T_{34} T_{45}) (T_{12} T_{23} T_{34}) (T_{12}).
\end{align}
This implementation requires the same $N-1 = 4$ looped beam-splitters as in the temporal implementation of Reck \textit{et al.}

\subsubsection{Triangular architecture due to de Guise~\textit{et al.}}

\textit{Decomposition.\textemdash} de Guise~\textit{et al.}\ in 2018~\cite{Guise2018} provided a recursive factorization of $U$ into $\text{U}(2)$ matrices.
The decomposition has a triangular structure similar to that of Reck~\textit{et al.} but is obtained by a different procedure that has appealing group-theoretic properties.

The decomposition proceeds recursively, and each step of the recursion proceeds by factorizing $n$-mode unitary matrices into a two-mode unitary matrix that is sandwiched between two $(n-1)$ mode unitary matrices~\cite{Murnaghan1952}.
Consider for example the case $N=5$.
The first round begins with factorizing $U_{5}$ according to
\begin{equation}
U_{5} = (\mathds{1}_{1} \oplus U_{4}) (U_{2} \oplus \mathds{1}_{3}) (\mathds{1}_{1} \oplus U_{4}),
\label{Eq:deGuiseFirstStep}
\end{equation}
where $\mathds{1}_{i}$ represents an $i\times i$ identity matrix.
Note that the $U_{2}$ matrix of \cref{Eq:deGuiseFirstStep} has a form of $T_{mn}$ and can be implemented using a single beam-splitter and a single phase-shifter.
The $U_{4}$ matrices are obtained by nulling all but the first element of the first row and column of $U_{5}$ as done in \cref{Eq:Reck0}.
The $T_{mn}$ matrices are then moved to the right hand side of the equality sign and regrouped to obtain $U_{2}$ and the second $U_{4}$ of \cref{Eq:deGuiseFirstStep}.

The first of the two $U_{4}$ matrices is then factorized into three-mode unitary matrices
\begin{align}
U_{5} = & (\mathds{1}_{2} \oplus U_{3}) (\mathds{1}_{1} \oplus U_{2} \oplus \mathds{1}_{2}) (\mathds{1}_{2} \oplus U_{3})\nonumber\\ 
&(U_{2} \oplus \mathds{1}_{3})(\mathds{1}_{1} \oplus U_{4}),
\end{align}
along the lines of the factorization of \cref{Eq:deGuiseFirstStep}.
This expression can be simplified by merging the second $U_{3}$ matrix into $U_{4}$, an operation that is allowed because the $(\mathds{1}_{2} \oplus U_{3})$ matrix commutes with the $(U_{2} \oplus \mathds{1}_{3})$ matrix on its right.
Thus,
\begin{align}
U_{5} = & (\mathds{1}_{2} \oplus U_{3}) (\mathds{1}_{1} \oplus U_{2} \oplus \mathds{1}_{2}) \cancel{(\mathds{1}_{2} \oplus U_{3})}\nonumber\\ 
&(U_{2} \oplus \mathds{1}_{3})(\mathds{1}_{1} \oplus U_{4}),\nonumber \\
 = & (\mathds{1}_{2} \oplus U_{3}) (\mathds{1}_{1} \oplus U_{2} \oplus \mathds{1}_{2}) (U_{2} \oplus \mathds{1}_{3})(\mathds{1}_{1} \oplus U_{4}).
 \label{Eq:deGuiseStepU3}
\end{align}
Finally, the remaining three-mode unitary, which is in the first factor of \cref{Eq:deGuiseStepU3}, is factorized into two-mode unitary matrices 
\begin{align}
U_{5} = & 
(\mathds{1}_{3} \oplus U_{2}) (\mathds{1}_{2} \oplus U_{2} \oplus \mathds{1}_{1}) (\mathds{1}_{3} \oplus U_{2}) \nonumber\\
& (\mathds{1}_{1} \oplus U_{2} \oplus \mathds{1}_{2}) (U_{2} \oplus \mathds{1}_{3})(\mathds{1}_{1} \oplus U_{4})
\end{align}
and one of the two-mode unitary matrices is absorbed by the four-mode unitary as 
\begin{align}
U_{5} = & 
(\mathds{1}_{3} \oplus U_{2}) (\mathds{1}_{2} \oplus U_{2} \oplus \mathds{1}_{1}) \cancel{(\mathds{1}_{3} \oplus U_{2})} \nonumber\\
& (\mathds{1}_{1} \oplus U_{2} \oplus \mathds{1}_{2}) (U_{2} \oplus \mathds{1}_{3})(\mathds{1}_{1} \oplus U_{4}).\nonumber\\
= & 
(\mathds{1}_{3} \oplus U_{2}) (\mathds{1}_{2} \oplus U_{2} \oplus \mathds{1}_{1}) (\mathds{1}_{1} \oplus U_{2} \oplus \mathds{1}_{2}) (U_{2} \oplus \mathds{1}_{3}) \nonumber\\
&(\mathds{1}_{1} \oplus U_{4}).
\end{align}
This factorization into two-mode unitary matrices and a single four-mode unitary completes one round of the decomposition.

The next round recursively factorizes $U_{4}$ into two mode unitary matrices and a three-mode unitary matrix.
The decomposition is complete when only $U_{2}$ unitary matrices remain.
This final decomposition has the form
\begin{align}
U_{5} = &(\mathds{1}_{3} \oplus U_{2}) (\mathds{1}_{2} \oplus U_{2} \oplus \mathds{1}_{1}) (\mathds{1}_{1} \oplus U_{2} \oplus \mathds{1}_{2}) (U_{2} \oplus \mathds{1}_{3}) \nonumber\\
&(\mathds{1}_{3} \oplus U_{2}) (\mathds{1}_{2} \oplus U_{2} \oplus \mathds{1}_{1}) (\mathds{1}_{1} \oplus U_{2} \oplus \mathds{1}_{2}) \nonumber\\
&(\mathds{1}_{3} \oplus U_{2}) (\mathds{1}_{2} \oplus U_{2} \oplus \mathds{1}_{1}) \nonumber\\
&(\mathds{1}_{3} \oplus U_{2})
\end{align}
The resulting decomposition has appealing group-theoretic properties such as allowing for an easy enumeration of Gelfan'd-Tseitlin patterns and enabling a straightforward recursive calculation of the Haar measure as presented in Ref.~\cite{Guise2018}.

\textit{Implementation.\textemdash}
This decomposition enables a spatial implementation with a structure similar to the Reck \textit{et al.}\ decomposition but with one difference.
In contrast to the Reck \textit{et al.} decomposition in which the $N$ extra phases are present in the diagonal unitary matrix $\mathds{D}_{5}$, here these phases are contained by $U_{2}$ matrices acting on the last two modes.
Thus, a spatial implementation of the de Guise \textit{et al.} decomposition differs from that of Reck \textit{et al.} in the location of the phase-shifters.
A temporal implementation of this decomposition is also possible using fast reconfigurable  phase-shifters and looped beam-splitters.

\subsection{Architectures based on decomposition into \texorpdfstring{$M$}{M}-mode unitary matrices}
\label{Sec:BackgroundUM}
Decompositions of $\text{SU}(N)$ transformations into $\text{U}(M)$ include the decompositions of Dhand and Goyal~\cite{Dhand2015} and of Su, Dhand \textit{et al.}~\cite{Su2019a}.

\subsubsection{CSD-based modular architecture of Dhand and Goyal}

\textit{Decomposition.\textemdash}
Ref.~\cite{Dhand2015} presented a decomposition with the motivation of realizing an arbitrary $N \times N$ unitary matrix $U_{N}$ in the spatial and internal degrees of freedom of light using (i.)~50-50 beam-splitters acting on spatial modes and (ii.)~optical elements that effect arbitrary transformations on the internal modes.
The decomposition presented in Ref.~\cite{Dhand2015} uses CSD to recursively decompose $U_{N}$.
The decomposition returns a sequence of $M \times M$ matrices the can be implemented as modules acting on $M$-dimensional internal modes of light such as polarization, time bins, temporal modes, and orbital angular momentum.
These arbitrary internal transformations, along with balanced (50-50) beam-splitters together realize $U_{N}$.

\begin{figure*}[htp!]
\subfloat[\label{Fig:SuCSDbased}]{
\includegraphics[width = 0.95\textwidth]{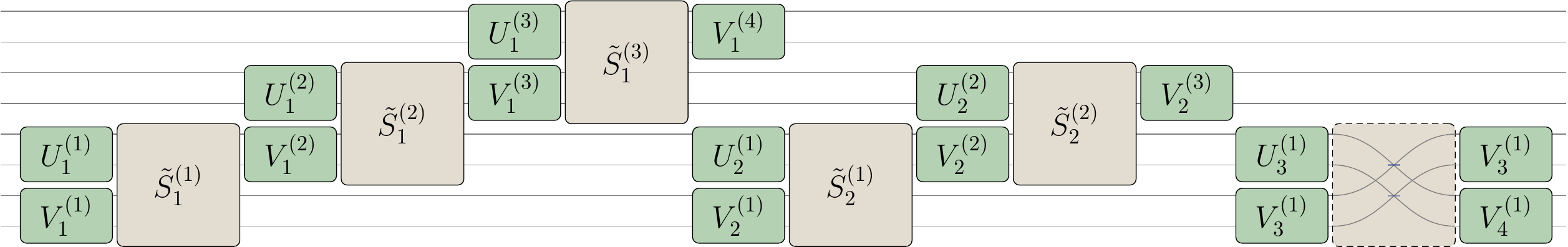}
}\\
\subfloat[\label{Fig:SuElimination}]{
\includegraphics[width=0.7\textwidth]{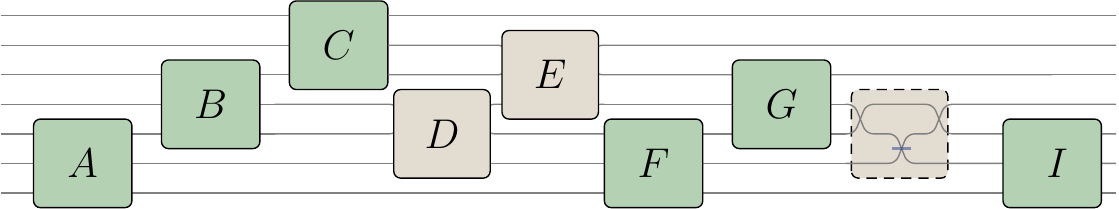}
}\caption{
(a) CS-based decomposition  of Ref.~\cite{Su2019a} of $N\times N$ unitary matrix into elementary matrices, including $M \times M$ universal unitary matrices (green) and specialized $2M \times 2M$ CS matrices (brown) for $N = 8$ and $M = 2$.
The subscript labels the layer that each unitary block belongs to, and the superscript distinguishes different unitary matrices within each layer.
Elimination-based decomposition of Ref.~\cite{Su2019a} for realizing an $\text{SU}(N)$ matrix using universal $\text{U}(M)$ and residual $\text{U}(2M-3)$ interferometers for $N, M = 7,3$.
The green and brown boxes represent universal and residual interferometers respectively.}
\end{figure*}

The decomposition of the given $U_{N}$ into smaller blocks proceeds as follows.
For concreteness, consider $N = 8$ and $M = 2$.
This could correspond to the realization of a $8\times 8$ unitary matrix on four spatial and two internal modes of light, for instance the two polarization modes of light. 
The first round begins with factorizing the given $U_{8}$ into
\begin{equation}
U_{8} = ({U}_{2} \oplus U_{6}) (\mathds{S}_{4} \oplus \mathds{1}_{4}) ({U}_{2} \oplus U_{6}),
\label{Eq:DhandGoyalFirst}
\end{equation}
using the CSD and setting CSD parameters $m = 2$ and $n = 6$.
Notice that \cref{Eq:DhandGoyalFirst} has a structure similar to that of~\cref{Eq:deGuiseFirstStep} of the de Guise \textit{et al.} decomposition.
The remaining decomposition procedure also shares similarities with the independently obtained decomposition of Ref.~\cite{Guise2018} described above.
The next step involves factorizing the $U_{6}$ matrix on the left using the CSD (setting $m = 2$ and $n = 4$) to obtain 
\begin{align}
U_{8} = &
(\mathds{1}_{2} \oplus {U}_{2} \oplus U_{4})
(\mathds{1}_{2} \oplus \mathds{S}_{4}\oplus \mathds{1}_{2})
(U_{2} \oplus U_{2} \oplus U_{4})\nonumber\\ 
&(\mathds{S}_{4} \oplus \mathds{1}_{4}) 
(\mathds{U}_{2} \oplus U_{6}),
\end{align}
and merging the second $U_{4}$ matrix into the $U_{6}$ transformation to obtain
\begin{align}
U_{8} = &
(\mathds{1}_{2} \oplus {U}_{2} \oplus U_{4})
(\mathds{1}_{2} \oplus \mathds{S}_{4}\oplus \mathds{1}_{2})
(U_{2} \oplus U_{2} \oplus \cancel{U_{4}})\nonumber\\ 
&(\mathds{S}_{4} \oplus \mathds{1}_{4}) 
(\mathds{U}_{2} \oplus U_{6}),\\
= &
(\mathds{1}_{2} \oplus {U}_{2} \oplus U_{4})
(\mathds{1}_{2} \oplus \mathds{S}_{4}\oplus \mathds{1}_{2})
(U_{2} \oplus U_{2} \oplus \mathds{1}_{4})\nonumber\\ 
&(\mathds{S}_{4} \oplus \mathds{1}_{4}) 
(\mathds{U}_{2} \oplus U_{6}).
\end{align}
More generally, in each step of the first round, a $U_{i}$ matrix (the final term of the first factor of the decomposition) is factorized using the CSD with $m = 2$ and $n = i - 2$ and the obtained matrices are merged into the $U_{N-M}$ matrix on their right whenever this is possible.
At the completion of the first round, only the $U_{6}$ matrix is left along with $U_{2}$ and $\mathds{S}_{4}$ matrices:
\begin{align}
U_{8} = &
(\mathds{1}_{4} \oplus U_{2} \oplus U_{2} )(\mathds{1}_{4} \oplus \mathds{S}_{4})\nonumber\\ 
 &
(\mathds{1}_{2} \oplus U_{2} \oplus U_{2} \oplus \mathds{1}_{2})(\mathds{1}_{2} \oplus \mathds{S}_{4}\oplus \mathds{1}_{2})\nonumber\\ 
 &
(U_{2} \oplus U_{2} \oplus \mathds{1}_{4})
(\mathds{S}_{4}\oplus \mathds{1}_{4})\nonumber\\ 
 &
(\mathds{U}_{2} \oplus U_{6}).
\label{Eq:FirstRound}
\end{align}
For ease of notation, let us define the first layer of matrices in the first three lines of \cref{Eq:FirstRound} into a single unitary $\mathds{V}_{8}$ to write
\begin{align}
U_{8} = 
(\mathds{V}_{8})
(\mathds{U}_{2} \oplus U_{6}).
\label{Eq:FirstRound2}
\end{align}

In the next round, the $U_{6}$ matrix is recursively factorized into a smaller $U_{4}$ matrix along with $U_{2}$ and $\mathds{S}_{4}$ according to
\begin{align}
U_{8} = &
(\mathds{V}_{8})
(\mathds{1}_{4} \oplus U_{2} \oplus U_{2} )(\mathds{1}_{4} \oplus \mathds{S}_{4})\nonumber\\ 
 &
(\mathds{1}_{2} \oplus U_{2} \oplus U_{2} \oplus \mathds{1}_{2})(\mathds{1}_{2} \oplus \mathds{S}_{4}\oplus \mathds{1}_{2})\nonumber\\ 
 &
(\mathds{U}_{2} \oplus U_{4}),\nonumber\\
 = &
(\mathds{V}_{8}) (\mathds{V}_{6})
(\mathds{U}_{2} \oplus U_{4}),
\label{Eq:DhandGoyalFinal}
\end{align}
where the newly introduced factors have a structure similar to the \cref{Eq:FirstRound} and are collected into the symbol $\mathds{V}_{6}$.
The decomposition ends with
\begin{align}
U_{8} =  (\mathds{V}_{8}) (\mathds{V}_{6}) (\mathds{V}_{4})(\mathds{V}_{2}),
\label{Eq:SuLabels}
\end{align}
which is depicted in \cref{Fig:SuCSDbased}, where different layers $\mathds{V}_{2j}$ are denoted by different subscripts.

Note that this decomposition has a triangular structure similar to that of the Reck \textit{et al.}\ decomposition and that of de Guise \textit{et al.}~\cite{Guise2018}.
In fact, the procedure gives identical beam-splitter parameters to that of Ref.~\cite{Guise2018} when $M$ is set to unity.

\textit{Implementation.\textemdash}
Ref.~\cite{Dhand2015} proposed an implementation of $\text{SU}(N)$ unitary matrices on an $M$-dimensional internal degree of freedom of light in $N/M$ spatial modes.
Each of the $\mathds{V}_{2j}$ matrices can be implemented using $2j-1$ internal transformations $U_{2}$ and $j-1$ CS matrices $\mathds{S}_{4}$, the latter of which can be implemented using two balanced beam-splitters and two internal transformations.
Examples of internal transformations include polarization, time bins, temporal modes and orbital angular momentum.
Internal transformations $U_{2}$ on the polarization basis can be realized using half and quarter waveplates~\cite{Simon1989,Simon1990}.
Universal linear optics on time bins can be performed using looped beam-splitters~\cite{Motes2014} and on temporal modes can be performed using quantum phase gates~\cite{Brecht2015a}.
Finally, universal operations on orbital angular momentum have been proposed in Ref.~\cite{Garcia-Escartin2011}.

\subsubsection{Modular architectures of Su, Dhand~\textit{et al.}}

\textit{Decomposition.\textemdash}
Ref.~\cite{Su2019a} recently presented architectures to implement arbitrary unitary transformations on multiple spatial and temporal modes of light using two approaches, firstly using a CSD based approach of Ref.~\cite{Dhand2015} and secondly using a systematic elimination-regrouping approach. 
These architectures allow combining the benefits of spatial architectures (namely low loss and parallel operation) with the scalability advantages of temporal architectures~\cite{Motes2014}. 

The elimination-regrouping based decomposition relies on systematically nulling elements of the given matrix using $T_{mn}$ matrices and regrouping the $T_{mn}$ matrices into $M$- and $(2M-3)$-mode unitary matrices.
Specifically, the given $\text{SU}(N)$ matrix is factorized into universal $\text{U}(M)$ matrices and specialized \textit{residual} unitary matrices acting on $(2M-3)$ modes.

As an illustration of the decomposition, consider $N = 7, M = 3$, i.e., a decomposition of a given $\text{SU}(7)$ matrix into universal $U_{3}$ matrices and specialized residual $\text{U}(3)$ matrices.
Consider an arbitrary $\text{SU}(7)$ matrix, which is denote by:
\begin{align}
\renewcommand{\arraystretch}{1.25} 
\left(\begin{array}{ccccccc}
*& C_{(1,2)}^{8}& C_{(2,3)}^{7}& B_{(3,4)}^{5}& B_{(4,5)}^{4}& A_{(5,6)}^{2}& A_{(6,7)}^{1}\\
&*& C_{(2,3)}^{9}& E_{(2,4)}^{11}& B_{(4,5)}^{6}& D_{(4,6)}^{10}& A_{(6,7)}^{3}\\
&& *& G_{(3,4)}^{16}& G_{(4,5)}^{15}& F_{(5,6)}^{13}& F_{(6,7)}^{12}\\
&&&*& G_{(4,5)}^{17}& H_{(4,6)}^{18}& F_{(6,7)}^{14}\\
&&&&*& I_{(5,6)}^{20}& I_{(6,7)}^{19}\\
&&&&&*& I_{(6,7)}^{21}\\
&&&&&&*
\end{array}\right),
\label{Eq:Su}
\end{align}
where only the elements above the diagonal are presented for simplicity.
The first step of the decomposition systematically nulls the entries labelled by the $A_{m,m}^{i}$ in the order $i$ using $T_{m,n}$ matrices according to
\begin{align}
U_{7} \, ({T}_{67} {T}_{56} {T}_{67})^{-1} = U_{\bar{A}},
\end{align}
where $U_{\bar{A}}$ is a matrix with zeros in place of the entries labelled by $A_{m,m}^{i}$ in \cref{Eq:Su}.
The motivation for using this ordering of nulling is that the terms in the bracket can be identified with universal $\text{U}(3)$ matrices acting on modes 5--7.
In the next two steps, the entries labelled  $B_{m,m}^{i}$ and $C_{m,m}^{i}$ are nulled:
\begin{equation*}
U_{7} \, ({T}_{67} {T}_{56} {T}_{67})^{-1}({T}_{45} {T}_{34} {T}_{45})^{-1} ({T}_{23} {T}_{12} {T}_{23})^{-1}
= U_{\bar{C}},
\end{equation*}
where all entries $A$ through $C$ have been nulled in $U_{\bar{C}}$.
The procedure now requires the nulling of the individual elements $D_{4,6}^{10}$ and $E_{2,4}^{11}$, which is performed using a non-nearest neighbor unitary matrices $T_{46}$ and $T_{24}$ according to
\begin{align}
U_{7} \, &({T}_{67} {T}_{56} {T}_{67})^{-1}({T}_{45} {T}_{34} {T}_{45})^{-1} \nonumber\\
&({T}_{23} {T}_{12} {T}_{23})^{-1}
{T}_{46}^{-1}{T}_{24}^{-1} 
= U_{\bar{E}}.
\end{align}
These $T_{24}$ and $T_{46}$ matrices are identified with specialized $U_{3}$ interferometers, which can be implemented using a single beam-splitter.
The decomposition concludes with
\begin{align}
{U_{7}} \, ({T}_{67} {T}_{56} {T}_{67})^{-1}
({T}_{45} {T}_{34} {T}_{45})^{-1}
({T}_{23} {T}_{12} {T}_{23})^{-1}
&\nonumber\\
{T}_{46}^{-1}{T}_{24}^{-1}
({T}_{67} {T}_{56} {T}_{67})^{-1}
({T}_{45} {T}_{34} {T}_{45})^{-1}
&\label{Eq:DecompositionU7}\\
{T}_{46}^{-1}
({T}_{67} {T}_{56} {T}_{67})^{-1}
&= {D}, \nonumber
\end{align}
or equivalently
\begin{align}
{U_{7}} =&  {D} ({T}_{67} {T}_{56} {T}_{67})
{T}_{46}
({T}_{45} {T}_{34} {T}_{45})
({T}_{67} {T}_{56} {T}_{67})
{T}_{24}
\nonumber\\
&
({T}_{23} {T}_{12} {T}_{23})
{T}_{46}
({T}_{45} {T}_{34} {T}_{45})
({T}_{67} {T}_{56} {T}_{67}).
\end{align}
These $T_{mn}$ can be grouped together into the following $\text{U}(3)$ matrices:
\begin{align}
{U_{7}} =&  \mathds{D}_{7} I_{567}H_{456}G_{345}F_{567}E_{234}D_{456}C_{123}B_{345}A_{567},
\end{align}
where the subscripts represent the indices of the modes that these matrices act on.
This obtained circuit is depicted in \cref{Fig:SuElimination}.
Observe that the structure of the full decomposition is also in a triangular form with many layers similar to \cref{Eq:SuLabels}.

\textit{Implementation.\textemdash}
The first, CSD based, implementation exploits the decomposition of~\cref{Eq:DhandGoyalFinal}.
It relies on identifying the $U_{M}$ and $\mathds{S}_{2M}$ matrices with universal and specialized linear interferometers acting on the spatial modes of light.
Each layer $\mathds{V}_{jM}$ of unitary matrices can be implemented on $jM$ pulses in $j$ temporal and $M$ spatial modes using only three reconfigurable interferometers that are coupled back to themselves using optical delay lines.

The implementation based on the elimination-regrouping decomposition relies on identifying both these types of matrices as transformations acting on the spatial modes of light and realizing these using reconfigurable linear interferometers.
Each of the layers is realized using two interferometers, whose outputs are fed back into their input using optical delay lines.
This completes an exposition of the existing architectures for linear optics. 

\section{Order of nulling for the CSD based decomposition}
\label{Sec:AppendixCSD}
Here we provide details about the order in which the matrix elements are nulled in the CSD based decomposition and the $m,n$ indices used to null these.
 add labels to the elements of the matrix $U_{12}$ to represent the order of nulling and the sequence of $m,n$ values used to null it.
Consider the matrix 
\begin{widetext}
\renewcommand{\arraystretch}{1.4} 
\begin{align*}
U_{12} = 
\left(\begin{array}{c|c|c|c|c|c|c|c|c|c|c|c}
*
&{\color{DarkViolet}D^{39,r}_{(01,02)}}
&{\color{DarkViolet}D^{38,r}_{(01,03)}}
&{\color{DarkViolet}D^{37,r}_{(01,04)}}
&{\color{DarkGreen}C^{36,\ell}_{(01,02)}}
&{\color{DarkGreen}C^{34,\ell}_{(01,03)}}
&{\color{DarkGreen}C^{31,\ell}_{(01,04)}}
&{\color{DarkRed}A^{05}_{(07,08)}}
&{\color{DarkRed}A^{04,r}_{(07,09)}}
&{\color{DarkRed}A^{03,r}_{(07,10)}}
&{\color{DarkRed}A^{02,r}_{(07,11)}}
&{\color{DarkRed}A^{01,r}_{(07,12)}}
\\
\hline
&*
&{\color{DarkViolet}D^{42,r}_{(02,03)}}
&{\color{DarkViolet}D^{41,r}_{(02,04)}}
&{\color{DarkViolet}D^{40,r}_{(02,05)}}
&{\color{DarkGreen}C^{35,\ell}_{(02,03)}}
&{\color{DarkGreen}C^{32,\ell}_{(02,04)}}
&{\color{DarkGreen}C^{29,\ell}_{(02,05)}}
&{\color{DarkRed}A^{09,r}_{(08,09)}}
&{\color{DarkRed}A^{08,r}_{(08,10)}}
&{\color{DarkRed}A^{07,r}_{(08,11)}}
&{\color{DarkRed}A^{06,r}_{(08,12)}}
\\
\hline
&
&*
&{\color{DarkViolet}D^{45,r}_{(03,04)}}
&{\color{DarkViolet}D^{44,r}_{(03,05)}}
&{\color{DarkViolet}D^{43,r}_{(03,06)}}
&{\color{DarkGreen}C^{33,\ell}_{(03,04)}}
&{\color{DarkGreen}C^{30,\ell}_{(03,05)}}
&{\color{DarkGreen}C^{28,\ell}_{(03,06)}}
&{\color{DarkRed}A^{12,r}_{(09,10)}}
&{\color{DarkRed}A^{11,r}_{(09,11)}}
&{\color{DarkRed}A^{10,r}_{(09,12)}}
\\
\hline
&
&
&*
&{\color{DarkViolet}D^{47,r}_{(04,05)}}
&{\color{DarkViolet}D^{46,r}_{(04,06)}}
&{\color{DarkGoldenrod}E^{49,r}_{(04,07)}}
&{\color{DarkBlue}B^{27,\ell}_{(04,05)}}
&{\color{DarkBlue}B^{25,\ell}_{(04,06)}}
&{\color{DarkBlue}B^{22,\ell}_{(04,07)}}
&{\color{DarkRed}A^{14,r}_{(10,11)}}
&{\color{DarkRed}A^{13,r}_{(10,12)}}
\\
\hline
&
&
&
&*
&{\color{DarkViolet}D^{48,r}_{(05,06)}}
&{\color{DarkGoldenrod}E^{51,r}_{(05,07)}}
&{\color{DarkGoldenrod}E^{50,r}_{(05,08)}}
&{\color{DarkBlue}B^{26,\ell}_{(05,06)}}
&{\color{DarkBlue}B^{23,\ell}_{(05,07)}}
&{\color{DarkBlue}B^{19,\ell}_{(05,08)}}
&{\color{DarkRed}A^{15,r}_{(11,12)}}
\\
\hline
&
&
&
&
&*
&{\color{DarkGoldenrod}E^{54,r}_{(06,07)}}
&{\color{DarkGoldenrod}E^{53,r}_{(06,08)}}
&{\color{DarkGoldenrod}E^{52,r}_{(06,09)}}
&{\color{DarkBlue}B^{24,\ell}_{(06,07)}}
&{\color{DarkBlue}B^{20,\ell}_{(06,08)}}
&{\color{DarkBlue}B^{16,\ell}_{(06,09)}}
\\
\hline
&
&
&
&
&
&*
&{\color{DarkGoldenrod}E^{56,r}_{(07,08)}}
&{\color{DarkGoldenrod}E^{55,r}_{(07,09)}}
&{\color{DarkCyan}F^{58,r}_{(07,10)}}
&{\color{DarkBlue}B^{21,\ell}_{(07,08)}}
&{\color{DarkBlue}B^{17,\ell}_{(07,09)}}
\\
\hline
&
&
&
&
&
&
&*
&{\color{DarkGoldenrod}E^{57,r}_{(08,09)}}
&{\color{DarkCyan}F^{60,r}_{(08,10)}}
&{\color{DarkCyan}F^{59,r}_{(08,11)}}
&{\color{DarkBlue}B^{18,\ell}_{(08,09,)}}
\\
\hline
&
&
&
&
&
&
&
&*
&{\color{DarkCyan}F^{63,r}_{(09,10)}}
&{\color{DarkCyan}F^{62,r}_{(09,11)}}
&{\color{DarkCyan}F^{61,r}_{(09,12)}}
\\
\hline
&
&
&
&
&
&
&
&
&*
&{\color{DarkCyan}F^{65,r}_{(10,11)}}
&{\color{DarkCyan}F^{64,r}_{(10,12)}}
\\
\hline
&
&
&
&
&
&
&
&
&
&*
&{\color{DarkCyan}F^{66,r}_{(11,12)}}
\\
\hline
&
&
&
&
&
&
&
&
&
&
&*
\end{array}\right).
\end{align*}
\end{widetext}
The elements of this matrix are nulled systematically in the order mentioned in the superscript.
The nulling is performed either from the left ($\ell$) using a $T_{mn}$ matrix or from the right (r) using a $T^{-1}_{mn}$ matrix.
The subscript refers to the $m,n$ values of the $T_{mn}$ or $T_{mn}^{-1}$ matrices that are used to null the specific element.
This completes the details of the CSD-based decomposition procedure.


%

\end{document}